\documentclass[]{jfm}
\usepackage{graphicx}
\usepackage{xcolor}
\usepackage{color}
\usepackage{soul}
\usepackage{newtxtext}
\usepackage{newtxmath}
\usepackage{amsmath, amssymb, amsfonts}
\usepackage{mathtools}
\usepackage{natbib}
\usepackage{hyperref}
\usepackage{physics}
\usepackage{mwe}
\usepackage{overpic}
\usepackage{placeins}
\usepackage{siunitx}
\usepackage{caption}
\usepackage{layouts}
\usepackage{indentfirst}

\graphicspath{{./pics/}}

\hypersetup{
    colorlinks = True,
    citecolor  = blue,
    linkcolor   = blue
}

\newcommand{\kB}{k_{\mathrm{B}}}

\newcommand{\Wi}{\mathrm{Wi}}
\newcommand{\Wisdk}{\mathrm{Wi}_{\mathrm{sd}}(k)}

\newcommand{\De}{\mathrm{De}}
\newcommand{\taup}{\tau_{\mathrm{p}}}

\newcommand{\tauf}{\tau_{\mathrm{f}}}
\newcommand{\taufk}{\tau_{\mathrm{f}}(k)}
\newcommand{\tauL}{\tau_{\mathrm{L}}}

\newcommand{\nuf}{\nu_{\mathrm{f}}}
\newcommand{\nup}{\nu_{\mathrm{p}}}

\newcommand{\Req}{R_{\mathrm{eq}}}
\newcommand{\Rmax}{R_{\mathrm{max}}}
\newcommand{\Lmax}{L_{\mathrm{max}}}
\newcommand{\Np}{N_{\mathrm{p}}}
\newcommand{\Npdash}{N_{\mathrm{p}}^{\prime}}

\newcommand{\taueta}{\tau_{\eta}}

\newcommand{\HL}{\color{Magenta}}

\usepackage{marginnote}
\usepackage[
    backgroundcolor=blue!15,
    bordercolor=blue,
    linecolor=blue,
    textsize=tiny,
]{todonotes}

\usepackage{changepage} 
\strictpagecheck        

\newcommand{\markerR}[2]{%
    \begingroup%
    \checkoddpage
    \setlength{\marginparwidth}{14mm}%
    \setlength{\marginparsep}{2mm}%
    \let\marginpar\marginnote%
    \ifoddpage%
        \normalmarginpar\todo[size=\tiny]{#1}
    \else%
        \reversemarginpar\todo[size=\tiny]{#1}
    \fi%
    \endgroup%
    {\HL #2}%
}

\strictpagecheck        
\newcommand{\markerL}[2]{%
    \begingroup%
    \checkoddpage
    \setlength{\marginparwidth}{14mm}%
    \setlength{\marginparsep}{7mm}%
    \let\marginpar\marginnote%
    \ifoddpage%
        \reversemarginpar\todo[
            size=\tiny,%
        ]{#1}%
    \else%
        \normalmarginpar\todo[
            size=\tiny,%
        ]{#1}%
    \fi%
    \endgroup%
    {\HL #2}%
}

\newcommand{\RomanNumeralCaps}[1]
\linenumbers

\title{Physical mechanism of turbulence attenuation by polymers from a timescale perspective}

\author{Hayato Masuda
    \corresp{\email{h\_masuda@fm.me.es.osaka-u.ac.jp}},
    Yusuke Koide,
    Yutaro Motoori
    \and Susumu Goto}

\affiliation{Graduate School of Engineering Science, The University of Osaka, 1-3 Machikaneyama, Toyonaka, Osaka, 560-8531, Japan}

\begin{document}
\maketitle

\begin{abstract}
    To elucidate the physical mechanism of turbulence attenuation by polymers at each scale, we conduct direct numerical simulations of homogeneous isotropic turbulence in dilute polymer solutions. We model polymers as FENE dumbbells and simulate them using Brownian dynamics. By visualising the hierarchical structures of coherent vortices, we demonstrate that as the Weissenberg number increases, polymers progressively suppress vortices from smaller to larger scales, attenuating their turbulent energy. While inspired by Lumley's theory, we propose a novel timescale-based framework for analysing this attenuation process using the scale decomposition. We define a scale-dependent Weissenberg number, $\Wi_{\mathrm{sd}}(k)$, as the ratio of the polymer relaxation time to the turnover time of multiscale vortices at each wave-number. We reveal that when expressed in terms of $\Wi_{\mathrm{sd}}(k)$, the energy attenuation rate at each scale collapses onto a single curve that rises at $\Wi_{\mathrm{sd}}(k) \gtrsim 1$, proving that $\Wi_{\mathrm{sd}}(k)$ successfully describes both the onset and the degree of turbulence attenuation at any given scale $k^{-1}$. Furthermore, the scale decomposition uncovers that polymers preferentially align with the turbulent stretching direction at the scale satisfying $\Wi_{\mathrm{sd}}(k) \approx 1$. Based on these results, we establish a physical picture of the polymer--turbulence interaction and explicitly link it to the statistics in turbulence attenuated by polymers.
\end{abstract}


\section{Introduction}
\label{sec:Introduction}
The addition of polymers to a liquid can significantly alter its flow state, even at dilute concentrations. For instance, adding a small amount of polymers to fully developed turbulence in a pipe drastically attenuates the turbulence. This phenomenon is widely known as the Toms effect \citep{Toms_1948}. As turbulence attenuation leads to a reduction in wall frictional drag, this effect has been applied to oil pipelines \citep{Burger_et_al_1982}, heat exchangers \citep{El-Azm_et_al_2021} and water discharge in firefighting \citep{Figueredo_and_Sabadini_2003}. In contrast, even in an extremely slow torsional shear flow, polymers induce a chaotic flow that exhibits several similarities to Newtonian turbulence \citep{Groisman_and_Steinberg_2000, Groisman_and_Steinberg_2001}. Thus, since polymers can modulate flow states over a broad range of Reynolds numbers, the underlying physical mechanisms have been extensively investigated for many years. In the present study, we focus on turbulence attenuation in dilute polymer solutions at high Reynolds numbers.

Turbulence attenuation by polymers arises from the mechanical interaction between polymers and flow fluctuations generated by the energy cascade \citep{White_and_Mungal_2008, Xi_2019, Zhang_et_al_2026}. In pioneering theoretical studies, \citet{Lumley_1969, Lumley_1973} proposed that this interaction occurs when the polymer relaxation time matches the turnover time of vortices. Subsequently, \citet{Tabor_and_de_Gennes_1986} and \citet{de_Gennes_1986} argued that the balance between the polymer elastic energy and the turbulent kinetic energy is also crucial for turbulence attenuation. These classical theories imply that both timescale and energy are important for the interaction between polymers and turbulence. However, these theories have not yet fully clarified the respective roles of timescale and energy in the physical mechanism of turbulence attenuation.


Based on these classical theories, turbulence attenuation has been investigated from both timescale and energy perspectives through experiments \citep{Friehe_and_Schwarz_1970, McComb_et_al_1977, Liberzon_et_al_2006, Ouellette_et_al_2009, Vonlanthen_and_Monkewitz_2013, Xi_et_al_2013, Zhang_et_al_2021}, numerical simulations \citep{Vaithianathan_and_Collins_2003, Terrapon_et_al_2004, Cai_et_al_2010, Perlekar_et_al_2010, de_Angelis_et_al_2013, Watanabe_and_Gotoh_2013, Valente_et_al_2014, Valente_et_al_2016, Rehman_et_al_2022, Rosti_et_al_2023, Garg_and_Rosti_2025, Chiarini_et_al_2026} and theoretical approaches \citep{Fouxon_and_Lebedev_2003, Xi_et_al_2013, Chiarini_et_al_2025}. In recent years, the understanding of turbulence attenuation from the energy perspective has deepened. For instance, to predict the critical length scale at which turbulence begins to attenuate, \citet{Xi_et_al_2013} proposed a theory based not on the energy balance suggested by \citet{Tabor_and_de_Gennes_1986} and \citet{de_Gennes_1986}, but on the energy flux balance between polymers and turbulence. Moreover, they validated this theory through their experiments on von K\'{a}rm\'{a}n turbulence in dilute polymer solutions. Additionally, experiments by \citet{Zhang_et_al_2021} revealed the existence of an elastic range at scales smaller than the inertial range but larger than the viscous dissipation range, where the second-order velocity structure function tends to follow $r^{1.38}$ (i.e.~the energy spectrum scales as $k^{-2.38}$). Motivated by this, \citet{Rosti_et_al_2023} conducted numerical simulations of homogeneous isotropic turbulence using the Oldroyd-B and FENE-P models, confirming the presence of this elastic range. Furthermore, through a spectral analysis of energy transfer, they demonstrated that the energy flux between polymers and turbulence is balanced at the critical length scale of this elastic range. These results successfully describe the macroscopic statistical properties of turbulence in dilute polymer solutions. However, previous studies have not yet fully explained the physical mechanisms underlying these statistical properties.

To advance our current understanding of turbulence attenuation, it is essential to clarify the dynamical processes of the interaction between polymers and vortices. Thus, the relevant timescales of polymers and vortices provide an important perspective, complementing the energy perspective. Recent experiments \citep{Zhang_and_Xi_2022} and numerical simulations \citep{Valente_et_al_2014, Valente_et_al_2016, Nguyen_et_al_2016, Rehman_et_al_2022} have investigated turbulence attenuation in terms of the ratio of the polymer relaxation time to the vortex turnover time at a single representative scale, such as the integral or Kolmogorov scale (i.e.~the Deborah number $\De$ or the Weissenberg number $\Wi$). However, no previous study has provided a framework for analysing turbulence attenuation using this timescale ratio at each scale. Consequently, the scale-by-scale interactions between polymers and multiscale vortices remain elusive. Such a framework is necessary because turbulence is composed of multiscale coherent vortices \citep{Goto_2008, Leung_et_al_2012, Goto_et_al_2017, Doan_et_al_2018, Motoori_and_Goto_2019, Motoori_and_Goto_2021, Fujino_et_al_2023, Goto_and_Motoori_2024}. \citet{Koide_and_Goto_2024} numerically demonstrated the effects of vortices at different scales on polymer dynamics by tracking the stretching and alignment of  polymers from a Lagrangian perspective under one-way coupling (i.e.~turbulence affects polymers without feedback from polymers). However, the scale-by-scale interactions under two-way coupling (i.e.~polymer feedback on the flow is included) remain unexplored.

Therefore, we aim to elucidate the physical mechanism of turbulence attenuation by polymers, providing a framework to quantify the scale-by-scale interactions between polymers and multiscale vortices. A key feature of the present study is that, building on Lumley's theoretical basis, we relate the polymer relaxation time to the characteristic timescale of turbulence at each scale. To this end, we employ a hybrid Eulerian--Lagrangian approach \citep{Watanabe_and_Gotoh_2010, Watanabe_and_Gotoh_2013} to conduct direct numerical simulations (DNS) of turbulence in dilute polymer solutions. This methodology couples DNS of turbulence with discrete element simulations of FENE dumbbells based on Brownian dynamics. By applying the polymer elastic stress to the fluid as a feedback force, we achieve two-way coupling between polymers and turbulence. Notably, this methodology does not require closure approximations for the constitutive law \citep{Peterlin_1966} or artificial diffusion for numerical stability \citep{Sureshkumar_and_Beris_1995, Dubief_et_al_2005}, both of which are typically inevitable in continuum polymer models based on the conformation tensor \citep{Bird_et_al_1980}. By varying the polymer relaxation time, we systematically investigate a wide range of Weissenberg numbers to thoroughly elucidate the timescale dependence of turbulence attenuation. Specifically, by employing scale decomposition to visualise vortices at individual scales, we demonstrate that polymers suppress small-scale vortices as the polymer relaxation time increases. For each scale, we further quantify the polymer--turbulence interaction by evaluating the attenuation rate of turbulent energy and the alignment of polymers with the stretching direction of the strain-rate field. Then, we organise these numerical results in terms of the scale-dependent Weissenberg number, defined as the ratio of the polymer relaxation time to the turnover time of vortices at each scale. This analysis clearly reveals the timescale-based physical mechanism of turbulence attenuation. Using this timescale-based understanding, we explain the dynamical processes underlying the behaviour of the turbulent energy spectrum in dilute polymer solutions.

In
\S\:\ref{sec:Method},
we provide an overview of the numerical simulations and describe the governing parameters. In
\S\S\:\ref{subsec:vortices_visualization}
and
\ref{subsec:energy_spectrum},
we qualitatively and quantitatively evaluate the turbulence attenuation caused by polymer additives. Subsequently, the physical mechanism of this attenuation is suggested from the perspective of timescales
(\S\:\ref{subsec:scale_local_weissenberg_number}).
Furthermore, we discuss the physical mechanism from a microscopic viewpoint by examining the polymer alignment with respect to the extensional direction of the turbulence
(\S\:\ref{subsec:polymer_alignment}).
In
\S\:\ref{sec:Discussion}, we summarise the physical picture underlying scale-by-scale turbulence attenuation by polymers (\S\:\ref{subsec:Physical_mechanism}) and link it to the spectral behaviour (\S\:\ref{subsec:spectral_behaviour}). We also discuss the scaling laws of the attenuated energy spectrum in the present study and previous findings (\S\:\ref{subsec:Discrepancy}).

\section{Method}
\label{sec:Method}
We conduct numerical simulations of homogeneous isotropic turbulence in dilute polymer
solutions. We use the Eulerian--Lagrangian approach proposed by
\citet{Watanabe_and_Gotoh_2013}. This method couples direct numerical
simulations (DNS) of turbulence with Brownian dynamics simulations (BDS) of
polymers. In the following, we describe the governing equations for the DNS
(\S~\ref{subsec:DNS_of_turbulence}) and the BDS
(\S~\ref{subsec:BDS_of_polymer}), and the control parameters of the system
(\S~\ref{subsec:Simulation_parameters}).

\subsection{Direct numerical simulations}
\label{subsec:DNS_of_turbulence}
The incompressible flow obeys the continuity equation
\begin{gather}
    \div \vb*{u} = 0 \label{eq:equation_of_continuity}
\end{gather}
and the Navier--Stokes equation
\begin{gather}
    \pdv{\vb*{u}}{t} + \vb*{u} \vb*{\cdot} \grad \vb*{u} = - \frac{1}{\rho} \grad p + \nuf \laplacian \vb*{u} + \vb*{f}_{\mathrm{p}} + \vb*{f}_{\mathrm{ex}}, \label{eq:NS_equation_of_polymer_solution}
\end{gather}
where $\vb*{u}(\vb*{x}, t)$ and $p(\vb*{x}, t)$ are the fluid
velocity and pressure at position $\vb*{x}$ and time $t$, $\rho$ is the density
of the solution, $\nuf$ is the kinematic viscosity of the solvent,
$\vb*{f}_{\mathrm{p}}$ is the force per unit mass exerted by the polymers on the
fluid, and $\vb*{f}_{\mathrm{ex}}$ is the external force driving homogeneous
isotropic turbulence. We numerically solve \eqref{eq:equation_of_continuity} and
\eqref{eq:NS_equation_of_polymer_solution} on a staggered grid using the
Simplified Marker and Cell (SMAC) method \citep{Harlow_and_Welch_1965}. For time
integration, we apply the second-order Crank--Nicolson method to the viscous
term, the first-order explicit Euler method to the pressure gradient term, and
the second-order Adams--Bashforth method to the remaining terms. We evaluate
spatial derivatives using the second-order central difference scheme.

To investigate whether the results are insensitive to the external
force, we employ two different forces. One is the force
$\vb*{f}_{\mathrm{ex}}^{(\mathrm{CI})}$, which provides a constant
energy injection rate
\citep{Lamorgese_and_Caughey_and_Pope_2004}. Its Fourier transform is given by
\begin{equation}
    \widehat{\vb*{f}}^{(\mathrm{CI})}_{\mathrm{ex}}(\vb*{k}) = \left\{
    \begin{array}{@{}c@{}l}
        \displaystyle \frac{P_{\mathrm{in}}}{2K_{\mathrm{low}}} \widehat{\vb*{u}}(\vb*{k}) & \ \ \ \  \qty(0 < \abs*{\vb*{k}} \leq k_{\mathrm{max}}), \\
        \vb*{0}                                                                            & \ \ \ \  \qty(\abs*{\vb*{k}} > k_{\mathrm{max}}),
    \end{array}
    \right. \label{eq:definition_of_external_force_inputrate_contant}
\end{equation}
where $\widehat{(\cdot)}$ denotes the Fourier transform of $(\cdot)$, $\vb*{k}$
is the wave-number vector, $P_{\mathrm{in}}$ is the energy injection rate,
$k_{\mathrm{max}}\:(= \sqrt{2})$ is the cutoff wave-number of the external
force, and $K_{\mathrm{low}}$ is the turbulent kinetic energy in the wave-number
range $0 < \abs{\vb*{k}} \leq k_{\mathrm{max}}$. The other is the random force
based on the Ornstein--Uhlenbeck (OU) process \citep{Eswaran_and_Pope_1988}. Its
Fourier transform is given by
\begin{equation}
    \widehat{\vb*{f}}^{(\mathrm{OU})}_{\mathrm{ex}}(\vb*{k}) = \left\{
    \begin{array}{@{}c@{}l}
        \displaystyle \widehat{\vb*{b}} - \qty(\vb*{k}\cdot \widehat{\vb*{b}})\frac{\vb*{k}}{\abs{\vb*{k}}^{2}} & \ \ \ \ \qty(0 < \abs{\vb*{k}} \leq k_{\mathrm{max}}), \\
        \vb*{0}                                                                                                 & \ \ \ \ \qty(\abs{\vb*{k}} > k_{\mathrm{max}}),
    \end{array}
    \right.
\end{equation}
where $\widehat{\vb*{b}}$ is a complex vector whose real and imaginary
parts independently follow the OU process satisfying
\begin{gather}
    \left\langle \widehat{b}_{i}\qty(\vb*{k}, t)\right\rangle = 0\ \ \ \ \mathrm{and}\ \ \ \ \left\langle \widehat{b}_{i}\qty(\vb*{k}, t)\widehat{b}_{j}^*\qty(\vb*{k}, t + s) \right\rangle = 2\sigma^2\delta_{ij}\exp(-\frac{s}{T_{\mathrm{L}}}). \label{eq:property_of_complex_vector}
\end{gather}
Here, $\langle \cdot \rangle$ denotes the ensemble average,
$(\cdot)^*$ denotes the complex conjugate, $\delta_{ij}$ is the Kronecker delta,
$\sigma$ is the amplitude, and $T_{\mathrm{L}}$ is the correlation time.

\subsection{Brownian dynamics simulations}
\label{subsec:BDS_of_polymer}
We model a polymer as a finitely extensible nonlinear elastic (FENE)
dumbbell \citep{Warner_1972}. The FENE dumbbell consists of two Brownian beads
connected by a nonlinear spring. The state of a polymer is represented by the
centre-of-mass position vector
\begin{gather}
    \vb*{r}_{\mathrm{g}} = \frac{\vb*{x}_{1} + \vb*{x}_{2}}{2} \label{eq:center_of_gravity_vector_of_polymer}
\end{gather}
and the end-to-end vector
\begin{gather}
    \vb*{R} = \vb*{x}_{1} - \vb*{x}_{2}, \label{eq:endtoend_vector_of_polymer}
\end{gather}
where $\vb*{x}_{1}$ and $\vb*{x}_{2}$ are the position vectors of the
beads. Assuming that each bead is subject to Stokes drag and that bead
inertia is negligible, we obtain the governing equations (Langevin equations)
for $\vb*{r}_{\mathrm{g}}$ and $\vb*{R}$ as
\begin{gather}
    \dv{\vb*{r}_{\mathrm{g}}}{t} = \vb*{u}(\vb*{r}_{\mathrm{g}}, t) + \frac{\Req}{\sqrt{8\taup}} \qty(\vb*{w}_{1} + \vb*{w}_{2}) \label{eq:langevin_equation_of_rg_vector}
\end{gather}
and
\begin{gather}
    \dv{\vb*{R}}{t} = \grad{\vb*{u}}(\vb*{r}_{\mathrm{g}}, t) \cdot
    \vb*{R} - \frac{1}{2\taup} \frac{\vb*{R}}{1 - \abs*{\vb*{R}}^2 /
        \Rmax^2} + \frac{\Req}{\sqrt{2\taup}} \qty(\vb*{w}_{1} -
    \vb*{w}_{2}), \label{eq:langevin_equation_of_R_vector}
\end{gather}
respectively. Here, $\Req$, $\Rmax$ and $\taup$ are the equilibrium
length, the maximum length and the relaxation time of the dumbbell,
respectively, and $\vb*{w}_{i}\:(i = 1 \text{ and } 2)$ are  Gaussian white noise processes satisfying
\begin{gather}
    \langle w_{i, j}\rangle = 0\ \ \ \ \mathrm{and}\ \ \ \ \langle w_{i, j}(t)w_{k, l}(t')\rangle = \delta_{ik}\delta_{jl}\delta\qty(t - t'),
\end{gather}
where $\delta (x)$ is the Dirac delta function. The relaxation time $\taup$
represents the characteristic time for the dumbbell to relax to its
equilibrium length $\Req$. We numerically integrate
\eqref{eq:langevin_equation_of_rg_vector} and
\eqref{eq:langevin_equation_of_R_vector} in time using the explicit Euler method
and the semi-implicit method \citep{Ottinger_1996}, respectively.

In \eqref{eq:NS_equation_of_polymer_solution}, we evaluate the force exerted by
polymers on the fluid as $\vb*{f}_{\mathrm{p}} = \div \vb*{T}$, where $\vb*{T}$ is
the polymer stress tensor. According to the formulation of
\citet{Watanabe_and_Gotoh_2013}, the components of the stress tensor $\vb*{T}$
are given by
\begin{equation}
    T_{ij} = \frac{\nup}{\taup} \qty(\frac{V}{\Np})\sum_{n = 1}^{\Np} \qty[\frac{1}{\Req^2}\frac{R^{(n)}_{i} R^{(n)}_{j}}{1 - \abs*{\vb*{R}^{(n)}}^2 / \Rmax^2} - \delta_{ij}] \delta_{\mathrm{weight}} \qty(\vb*{x} - \vb*{r}_{\mathrm{g}}^{(n)}),  \label{eq:polymer_stress}
\end{equation}
where $(\cdot)^{(n)}$ denotes a quantity for the $n$-th polymer, $\nup$ is
the polymer contribution to the kinematic viscosity, $N_{\mathrm{p}}$ is the total number of
polymers, $V\:(= L_{\mathrm{box}}^3)$ is the volume of the computational domain,
and $\delta_{\mathrm{weight}}(\vb*{x})$ is the weight function defined with the
DNS grid spacing $\Delta$ as
\begin{equation}
    \delta_{\mathrm{weight}}\qty(\vb*{x}) = \left\{
    \begin{array}{@{}c@{}l}
        \displaystyle \frac{1}{\Delta^3}\qty(1 - \frac{\abs{x}}{\Delta})\qty(1 - \frac{\abs{y}}{\Delta})\qty(1 - \frac{\abs{z}}{\Delta}) & \ \ \ \  \qty(\abs{\vb*{x}} \leq \Delta), \\
        0                                                                                                                                & \ \ \ \  \qty(\abs{\vb*{x}} > \Delta).
    \end{array}
    \right.
    \label{eq:polymer_stress_weight}
\end{equation}
In addition, using the Stokes drag coefficient $\zeta$, spring
constant $h$, Boltzmann constant $\kB$ and temperature $T$, the
parameters $\taup$ and $R_{\mathrm{eq}}$ are related as $\taup =
    \zeta / (4h)$ and $\Req = \sqrt{\kB T / h}$, respectively.

\subsection{Parameters}
\label{subsec:Simulation_parameters}
\begin{table}
    \begin{center}
        \begin{tabular}{c c c c c c c c c c c c }
            $\vb*{f}_{\mathrm{ex}}$                 & $L_{\mathrm{box}}$ & $N_{\mathrm{grid}}$ & $k_{\mathrm{max}}$ & $\nu_{\mathrm{f}}$ & $P_{\mathrm{in}}$  & $\epsilon$         & $L$   & $\eta$             & $\tau_{\mathrm{L}}$ & $\tau_{\eta}$ & $\Re_{\lambda}$             \\ \hline
            $\vb*{f}_{\mathrm{ex}}^{\mathrm{(CI)}}$ & $2\pi$             & $512^3$             & $\sqrt{2}$         & $1.0\times10^{-3}$ & $4.4\times10^{-2}$ & $4.4\times10^{-2}$ & $1.1$ & $1.2\times10^{-2}$ & $2.1$               & $0.15$        & $150$          \vspace{2mm} \\
            $\vb*{f}_{\mathrm{ex}}^{\mathrm{(OU)}}$ & $2\pi$             & $512^3$             & $\sqrt{2}$         & $1.0\times10^{-3}$ & $7.3\times10^{-2}$ & $7.4\times10^{-2}$ & $1.0$ & $1.1\times10^{-2}$ & $1.7$               & $0.12$        & $167$
        \end{tabular}
        \caption{
            Parameters of the Newtonian turbulent flow field used as the initial condition
            for the DNS. Here, $L_{\mathrm{box}}$ is the side length of the computational
            domain, $N_{\mathrm{grid}}$ is the number of grid points, $k_{\mathrm{max}}$ is
            the cutoff wave-number of the external force, $\nuf$ is the kinematic viscosity
            of the solvent, $P_{\mathrm{in}}$ is the energy injection rate, $\epsilon$ is
            the energy dissipation rate, $L$ is the integral length, $\eta$ is the
            Kolmogorov length, $\tauL$ is the integral time, $\taueta$ is the
            Kolmogorov time, and $\Re_{\lambda}$ is the Taylor microscale Reynolds
            number.
        }
        \label{tab:dns_parameters}
    \end{center}
\end{table}

The initial condition for the fluid is a Newtonian turbulent flow.
We summarise the parameters of this flow in table~\ref{tab:dns_parameters}. The
Taylor
microscale Reynolds number is $\Re_{\lambda} \approx 150$ for both external
forces. We uniformly and isotropically
disperse FENE dumbbells with the initial length of $R\:(=\abs{\vb*{R}}) = \Req$. Assuming
that the dynamics of FENE dumbbells within the same computational grid cell are
statistically equivalent, we simulate a reduced number of polymers,
$N'_{\mathrm{p}}\:(= N_{\mathrm{p}} / C_{\mathrm{rep}})$, where $N_{\mathrm{p}}$
is the actual number of polymers and $C_{\mathrm{rep}}$ is the replica
parameter. In the present study, we set $N'_{\mathrm{p}} / N_{\mathrm{grid}} =
    10$, where $N_{\mathrm{grid}}\:(= 512^3)$ is the total number of grid points.
We confirm in
appendix~\ref{sec:APPENDIX_dependence_of_the_number_of_polymers}
that when $N'_{\mathrm{p}} \gtrsim 10$, the turbulent energy
spectrum is insensitive to $N'_{\mathrm{p}}$ over the wave-number range except for the viscous dissipation range.

We consider a polymer solution characterised by the concentration, maximum length $R_{\mathrm{max}}$ and relaxation time $\taup$.
In the present study, we fix the viscosity ratio of the polymer
contribution to the solvent viscosity
\begin{equation}
    \xi = \frac{\nup}{\nuf}
\end{equation}
and the ratio of the maximum length to the equilibrium length
\begin{equation}
    \Lmax = \frac{\Rmax}{\Req},
\end{equation}
while we change $\taup$. Here, $\xi$ is proportional to the polymer
concentration under the same $R_{\mathrm{max}}$ and
$\taup$. We consider the case where the polymers can store a large
amount of elastic energy. Therefore, we choose large values of
$\xi\:(= 0.1)$ and $\Lmax\:(= 100)$. With this choice, the elastic
energy of stretched polymers becomes comparable to the turbulent
kinetic energy, as shown in appendix \ref{sec:APPENDIX_energy}. Then,
we systematically change the Weissenberg number,
\begin{equation}
    \Wi = \frac{\taup}{\taueta}, \label{eq:weissenberg_number}
\end{equation}
where $\taueta$ is the Kolmogorov time of the Newtonian turbulence. In the
present study, we investigate a wide range of $\Wi$ from $0.06$ to $1024$. This
range corresponds to the Deborah number, $\De\:(= \taup / \tauL)$, from $0.004$ to $68$,
where $\tauL$ is the integral time. Thus, $\taup$ ranges from times
shorter than the Kolmogorov time to those longer than the integral
time. For both types of forcing, $\De = 1$ corresponds to $\Wi
    \approx 14$.

\section{Turbulence attenuation}
\label{sec:Turbulence_attenuation}
\begin{figure}
    \begin{center}
        \begin{minipage}{.49\linewidth}
            \centering
            \begin{overpic}[width = 1.7in]{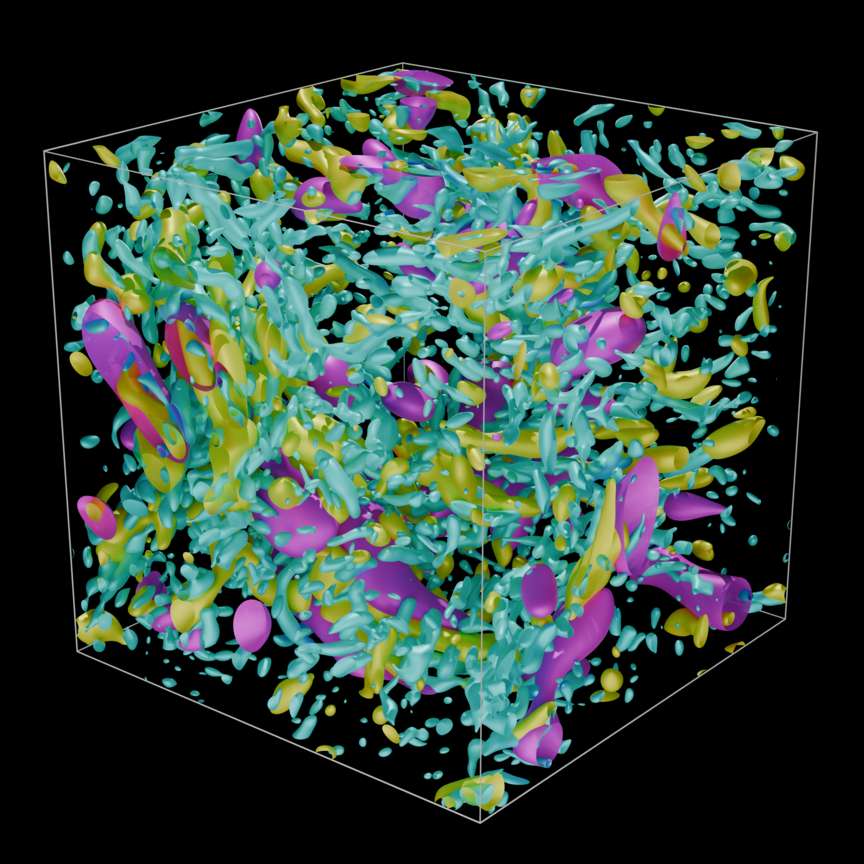}
                \put(3, 90){\textcolor{white}{\large($a$)}}
            \end{overpic}
        \end{minipage}
        \hspace*{-21.7mm}   
        \begin{minipage}{.49\linewidth}
            \centering
            \begin{overpic}[width = 1.7in]{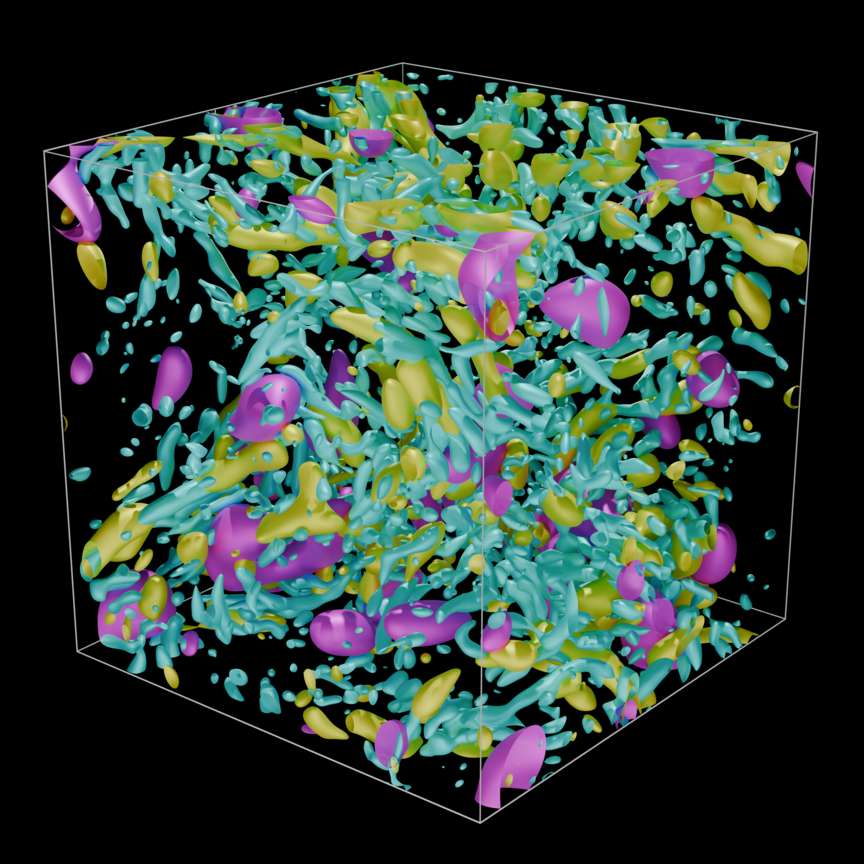}
                \put(3, 90){\textcolor{white}{\large($b$)}}
            \end{overpic}
        \end{minipage} \\
        \vspace*{3mm}
        \begin{minipage}{.49\linewidth}
            \centering
            \begin{overpic}[width = 1.7in]{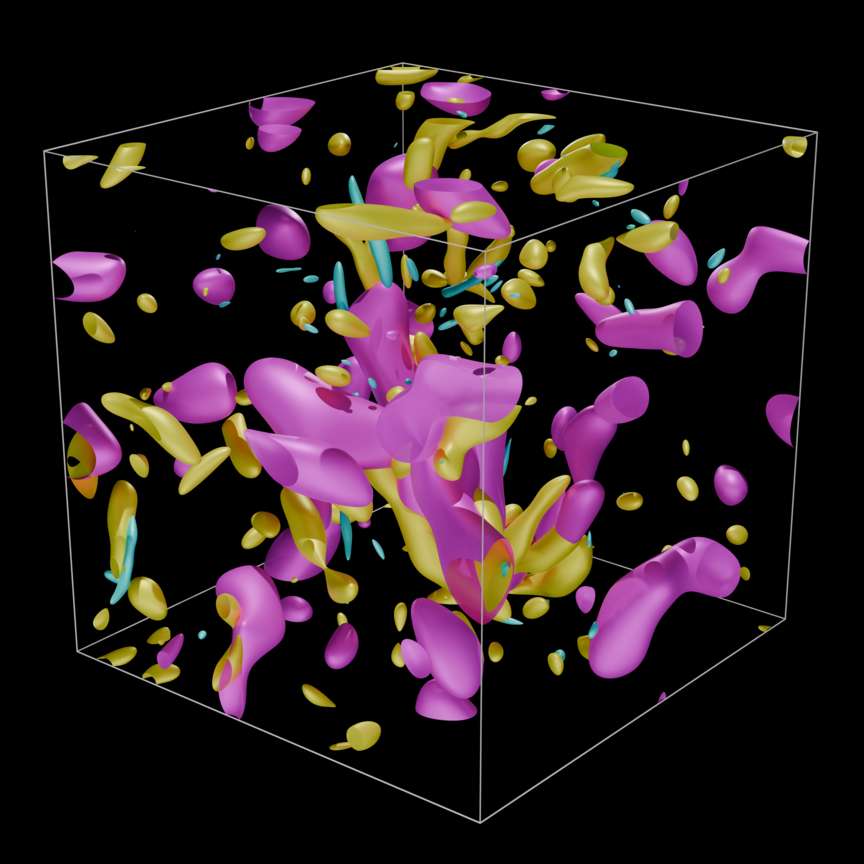}
                \put(3, 90){\textcolor{white}{\large($c$)}}
            \end{overpic}
        \end{minipage}
        \hspace*{-21.7mm}   
        \begin{minipage}{.49\linewidth}
            \centering
            \begin{overpic}[width = 1.7in]{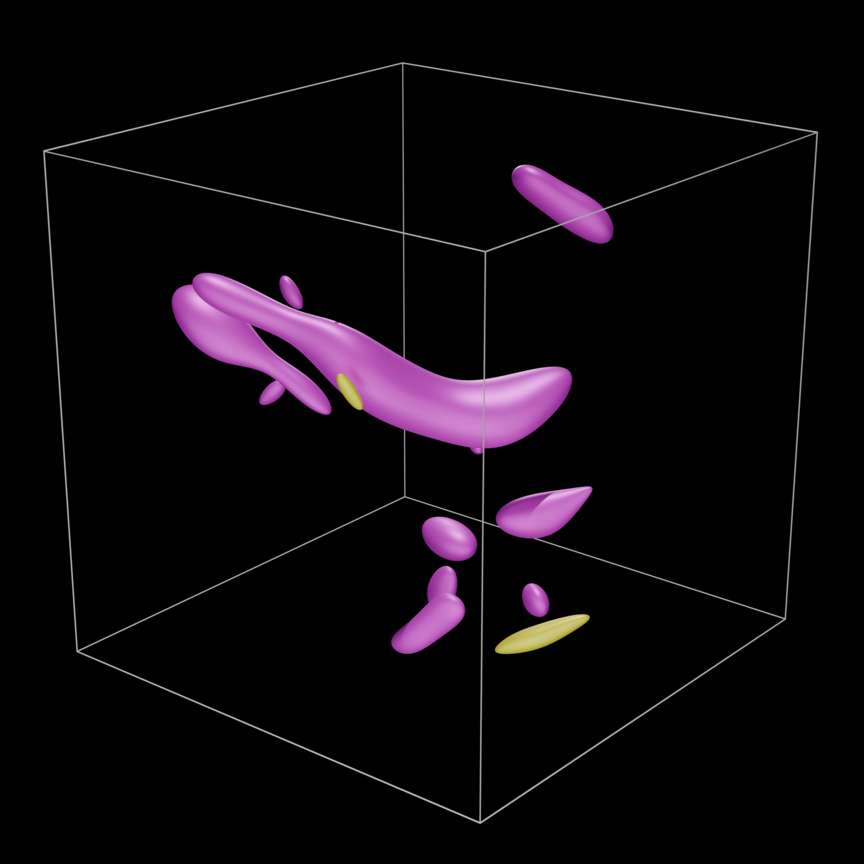}
                \put(3, 90){\textcolor{white}{\large($d$)}}
                \put(4.25, 15.5){\includegraphics[width=1cm, interpolate=false]{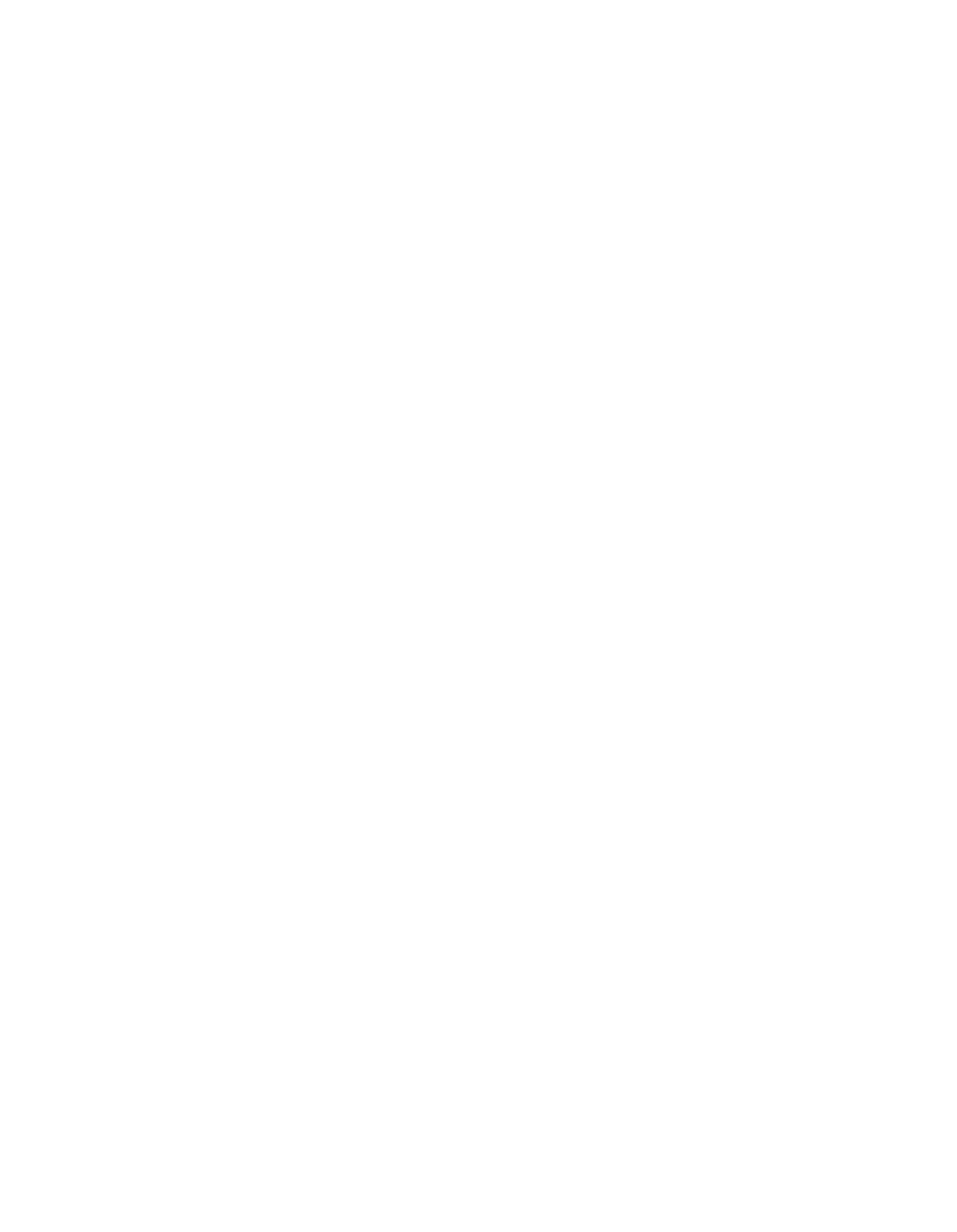}}
            \end{overpic}
        \end{minipage}
        \caption{
            Positive isosurfaces of the second invariant $Q^{(k_{\mathrm{c}})}$ of the velocity gradient
            tensor at three different scales $k_{\mathrm{c}}\eta = 0.06$ (magenta),
            $0.12$ (yellow) and $0.24$ (cyan) for ($a$) the Newtonian turbulence and the
            polymer solutions at ($b$) $\Wi = 0.06$, ($c$) $8$ and ($d$) $32$.  The
            thresholds are $Q^{(k_{\mathrm{c}})} = 0.88$ (magenta), $3$ (yellow) and $8$ (cyan),
            which are twice the standard deviation of $Q^{(k_{\mathrm{c}})}$ at the corresponding scale
            in the Newtonian turbulence.
        }
        \label{fig:vorticies_visualization}
    \end{center}
\end{figure}
\subsection{Visualisation of coherent vortices}
\label{subsec:vortices_visualization}
First, we qualitatively demonstrate the turbulence attenuation by
polymers in the physical space.
Figure~\ref{fig:vorticies_visualization} visualises vortices in the
statistically steady state ($t /\tau_{\mathrm{L}} \gtrsim 10$) of the polymer-laden turbulence driven by
$\vb*{f}_{\mathrm{ex}}^{(\mathrm{CI})}$. These vortices are
identified by positive isosurfaces of the second invariant $Q^{(k_{\mathrm{c}})}$
of the velocity gradient tensor, which is evaluated from the
band-pass filtered velocity field. This filter passes only the
wave-number range $k_{\mathrm{c}} / 2 < \abs{\vb*{k}} \leq
    k_{\mathrm{c}}$, where $k_{\mathrm{c}}$ is the cutoff wave-number.
We set $k_{\mathrm{c}}\eta = 0.06$ (magenta),
$0.12$ (yellow) and $0.24$ (cyan). We fix the threshold of
$Q^{(k_{\mathrm{c}})}$ at twice the standard deviation of $Q^{(k_{\mathrm{c}})}$ in the
Newtonian turbulence. In the case without polymers
    [figure~\ref{fig:vorticies_visualization}($a$)], i.e.~the Newtonian
turbulence, vortices of all three scales are well developed.
Figures~\ref{fig:vorticies_visualization}($b$--$d$) show results for
$\Wi = 0.06$, $8$ and $32$.  For $\Wi < 1$
[figure~\ref{fig:vorticies_visualization}($b$)], we see that
hierarchical structures of vortices are also developed similarly to
those in the Newtonian turbulence. In contrast, when $\Wi$ is larger
than unity [figure~\ref{fig:vorticies_visualization}($c$,$d$)],
vortices tend to be attenuated. We also see that this attenuation
starts from smaller-scale vortices as $\Wi$ increases.

\begin{figure}
    \begin{center}
        \begin{minipage}{0.329\linewidth}
            \begin{overpic}[width = 1.7in]{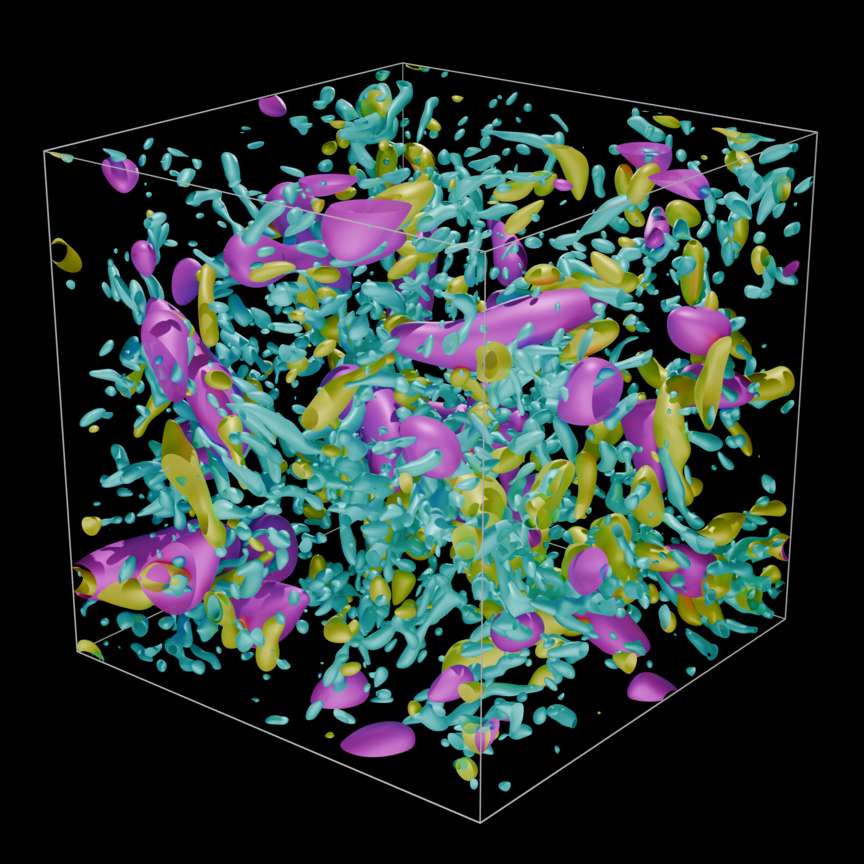}
                \put(3, 90){\textcolor{white}{\large($a$)}}
            \end{overpic}
        \end{minipage}  
        \begin{minipage}{0.329\linewidth}
            \begin{overpic}[width = 1.7in]{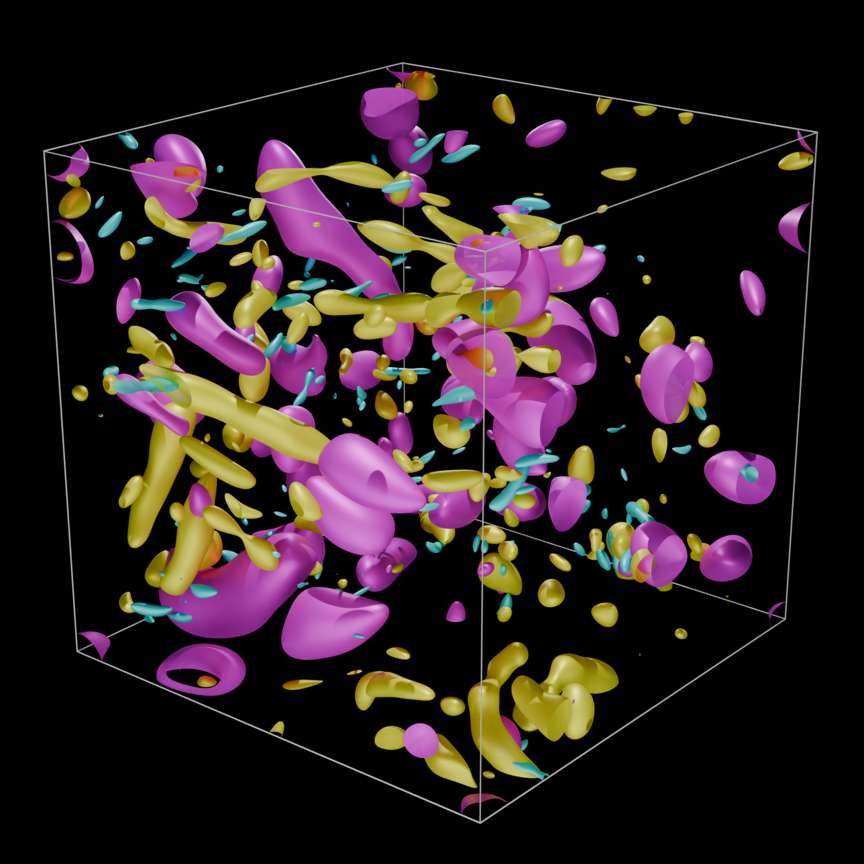}
                \put(3, 90){\textcolor{white}{\large($b$)}}
            \end{overpic}
        \end{minipage}
        \begin{minipage}{0.329\linewidth}
            \begin{overpic}[width = 1.7in]{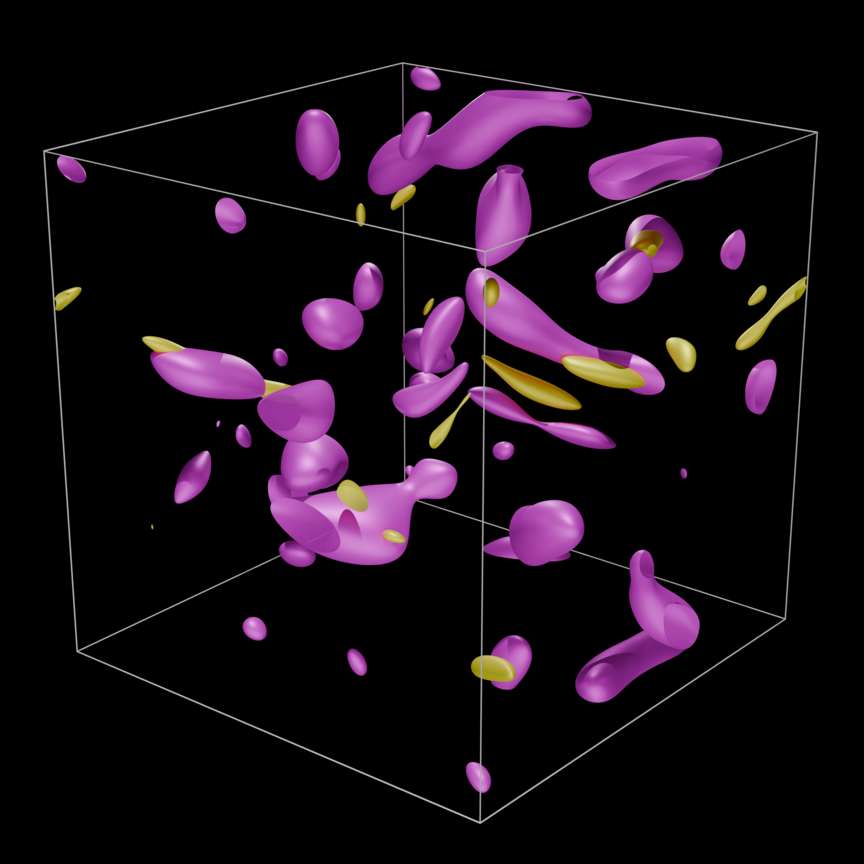}
                \put(3, 90){\textcolor{white}{\large($c$)}}
                \put(4.25, 15.5){\includegraphics[width=1cm, interpolate=false]{axis.pdf}}
            \end{overpic}
        \end{minipage}
        \caption{
            Positive isosurfaces of the second invariant $Q^{(k_{\mathrm{c}})}$ of the velocity gradient
            tensor at three different scales $k_{\mathrm{c}}\eta = 0.06$ (magenta),
            $0.12$ (yellow) and $0.24$ (cyan) for the polymer solution at $\Wi = 32$ at
            times ($a$) $t / \tau_{\eta} = 0$, ($b$) $20$ and ($c$) $45$ after polymer
            addition. The thresholds are the same as those in
            figure~\ref{fig:vorticies_visualization}.
        }
        \label{fig:temporal_evolution_of_turbulence_attenuation}
    \end{center}
\end{figure}

Next, we examine how turbulence attenuation develops in time after polymers are
added to the Newtonian turbulence. We visualise in
figure~\ref{fig:temporal_evolution_of_turbulence_attenuation} vortices at the
three scales for the polymer solution at $\Wi =
    32$ before the flow reaches the statistically steady state
    [figure~\ref{fig:vorticies_visualization}($d$)]. Panels show different times
($a$) $t / \tau_{\eta} = 0$, ($b$) $20$ and ($c$) $45$ after polymer addition.
We see that small-scale vortices (cyan) are first attenuated
    [figure~\ref{fig:temporal_evolution_of_turbulence_attenuation}($b$)], and
afterwards medium-scale vortices (yellow) are attenuated
    [figure~\ref{fig:temporal_evolution_of_turbulence_attenuation}($c$)]. This
result implies that polymers with $\Wi \gtrsim 1$ attenuate vortices
successively from smaller to larger scales.

\begin{figure}
    \begin{center}
        \begin{overpic}[tics=10]{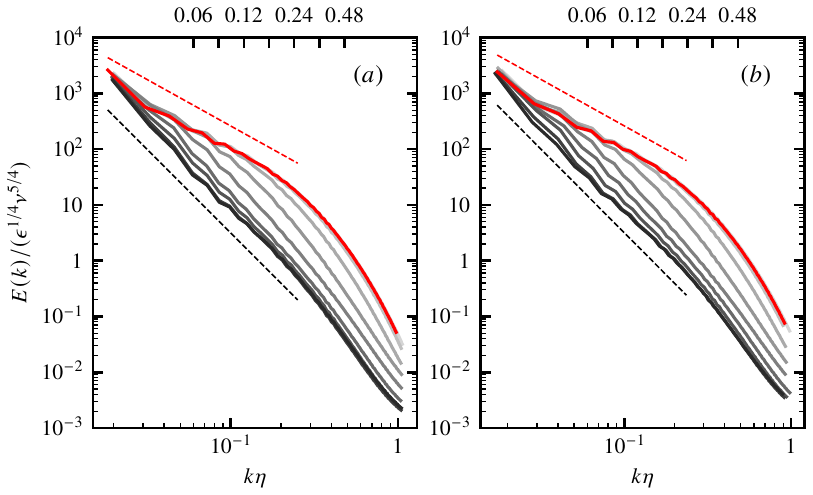}
        \end{overpic}
        \caption{
        Average energy spectrum $E(k)$ of turbulence driven by the external forces
        ($a$) $\vb*{f}_{\mathrm{ex}}^{(\mathrm{CI})}$ and ($b$)
        $\vb*{f}_{\mathrm{ex}}^{(\mathrm{OU})}$. The red line indicates the result for
        the Newtonian turbulence. The grey lines, from lighter to darker, correspond to $\Wi
            = 0.06$, $0.25$, $0.5$, $1$, $2$, $4$, $8$, $16$, $32$, $64$, $256$ and $1024$.
        The red and black dashed lines show $E(k) \propto k^{-5/3}$ and $k^{-3}$,
        respectively.}
        \label{fig:energy_spectrum}
    \end{center}
\end{figure}


\subsection{Energy spectrum}
\label{subsec:energy_spectrum}
To quantitatively show the turbulence attenuation, we plot in
figure~\ref{fig:energy_spectrum} the time-averaged energy spectrum $E(k)$
in the statistically steady state ($t / \tau_{\mathrm{L}} \gtrsim 10$) for turbulence driven by ($a$)
$\vb*{f}_{\mathrm{ex}}^{(\mathrm{CI})}$ and ($b$)
$\vb*{f}_{\mathrm{ex}}^{(\mathrm{OU})}$.
The red line indicates the result for the Newtonian turbulence. The darker grey
lines indicate the results for larger $\Wi$.  We
normalise $k$ and $E(k)$ by the kinematic viscosity $\nu\:(=\nuf + \nup)$ and
energy dissipation rate $\epsilon\:(=\epsilon_{\mathrm{f}} +
    \epsilon_{\mathrm{p}})$ of the solution, where $\epsilon_{\mathrm{f}}$ and
$\epsilon_{\mathrm{p}}\:(=\overline{\vb*{u} \cdot \vb*{f}_{\mathrm{p}}})$
are the average energy dissipation rates due to turbulence and polymers,
respectively, and $\eta\:(=\epsilon^{-1/4}\nu^{3/4})$ is the Kolmogorov length
of the solution. Here, $\overline{(\cdot)}$ denotes the time and spatial average. We see that in
the case with $\Wi \lesssim 1$ (lighter lines), $E(k)$ is almost
identical to that for the Newtonian turbulence. In contrast, for
larger $\Wi$ (darker lines), we see that the attenuation of $E(k)$ occurs first at
higher wave-numbers and then extends to lower wave-numbers. This
attenuation is due to the coil-stretch transition of the polymers
for $\Wi \gtrsim \mathcal{O}(1)$, which causes energy transfer from
turbulence to polymers \citep{Watanabe_and_Gotoh_2010}. We will discuss this mechanism in detail in \S\:\ref{subsec:polymer_alignment}.

We can also see that the energy spectrum changes from $k^{-5/3}$
(the red dashed line) to $k^{-3}$ (the black dashed one) scalings as
$\Wi$ increases. More specifically, for the Newtonian turbulence and
polymer solutions with $\Wi \lesssim 1$, the energy spectrum follows
$E(k) \propto k^{-5/3}$ in the wave-number range $0.06 \lesssim
    k\eta \lesssim 0.24$; see the upper axis of
figure~\ref{fig:energy_spectrum}. In contrast, the spectrum tends to
follow $E(k) \propto k^{-3}$ in the wave-number range where
turbulence is attenuated.
Although this tendency is more significant for larger $\Wi$, as $\Wi$ increases, it
begins to saturate at $\Wi \approx 16$, and then almost saturates
for $\Wi \gtrsim 256$. In these saturated cases, the spectrum
also follows $E(k) \propto k^{-3}$ for $0.06 \lesssim k\eta \lesssim
    0.24$. For higher wave-numbers, the spectrum becomes steeper than
$k^{-3}$, because the effect of viscosity becomes non-negligible.
This scaling $k^{-3}$ implies that the timescales of attenuated
flow become comparable to the polymer relaxation time $\taup$. We
discuss the physics behind these spectral behaviour in \S\:\ref{subsec:spectral_behaviour}
and its relation to the previous studies in \S\:\ref{subsec:Discrepancy}.

\begin{figure}
    \begin{center}
        \begin{overpic}[tics=10]{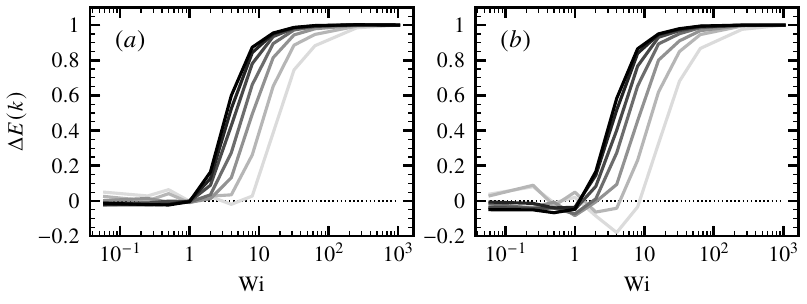}
        \end{overpic}
        \caption{
        Relative attenuation rate $\Delta E(k)$ of the energy spectrum as a function of
        $\Wi$ for turbulence driven by the external forces ($a$)
        $\vb*{f}_{\mathrm{ex}}^{(\mathrm{CI})}$ and ($b$)
        $\vb*{f}_{\mathrm{ex}}^{(\mathrm{OU})}$. The lines, from lighter to darker,
        correspond to the wave-numbers $k \eta = 0.06$, $0.085$, $0.12$, $0.17$, $0.24$,
        $0.34$ and $0.48$.
        }
        \label{fig:energy_attenuation_rate_afo_Wi}
    \end{center}
\end{figure}

Next, we examine how the turbulence attenuation depends on scale.  To this
end, we define the relative attenuation rate of the energy spectrum $E(k)$ at a wave-number $k$ as
\begin{equation}
    \Delta E(k) = \frac{E^{(\mathrm{N})}(k) - E(k)}{E^{(\mathrm{N})}(k) - E_{\infty}(k)},
    \label{eq:definition_of_energy_attenuation_rate}
\end{equation}
where $E^{(\mathrm{N})}$ is the energy spectrum for the Newtonian turbulence, and
$E_{\infty}$ is the energy spectrum for $\Wi = 1024$ where the attenuation
saturates. Thus, $\Delta E(k)\:(> 0)$ represents the relative attenuation rate of
turbulent energy at the scale $k^{-1}$, and $\Delta E(k) = 1$ indicates the
saturation of turbulence attenuation.

Figure~\ref{fig:energy_attenuation_rate_afo_Wi} shows $\Delta E(k)$ as a
function of $\Wi$ for turbulence driven by ($a$)
$\vb*{f}_{\mathrm{ex}}^{(\mathrm{CI})}$ and ($b$)
$\vb*{f}_{\mathrm{ex}}^{(\mathrm{OU})}$. The lines, from lighter to darker
colour, correspond to the wave-numbers $k \eta = 0.06$, $0.085$, $0.12$, $0.17$,
$0.24$, $0.34$ and $0.48$. We see that $\Delta E(k)$
shows similar behaviour for the two forcing methods.  For $\Wi
    \lesssim 1$ (i.e.~$\taup \lesssim \tau_{\eta}$), we see $\Delta E(k) \approx 0$
irrespective of the wave-number $k$. In other words, polymers with
$\tau_{\mathrm{p}} \lesssim \tau_{\eta}$ do not attenuate the turbulent energy.
In contrast, for $\Wi \approx 1$, the turbulence attenuation first appears at
the smallest scale (i.e.~$k\eta=0.48$ indicated by the darkest
line), and for larger $\Wi$, it extends to larger scales (indicated
by the lighter lines). For $\Wi \gtrsim 10$, the attenuation almost
saturates at the smallest scale, and for $\Wi \gtrsim 10^{2}$, the
attenuation saturates at all scales.

\begin{figure}
    \begin{center}
        \includegraphics[width = 0.5\linewidth]{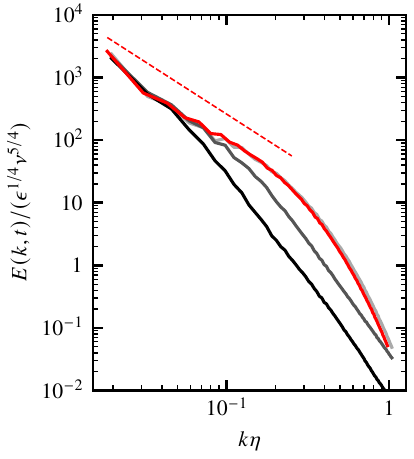}
        \caption{
            Temporal evolution of the energy spectrum $E(k, t)$ during the transient attenuation process for
            the polymer solution at $\Wi = 32$.  The red line
            indicates the result for the Newtonian turbulence. The
            grey lines, from lighter to darker, correspond to $t /
                \tau_{\eta} = 0$, $20$ and $45$, respectively,
            which are the same times as those in
            figure~\ref{fig:temporal_evolution_of_turbulence_attenuation}.
            The red dashed line shows $E(k) \propto k^{-5/3}$.
        }
        \label{fig:temporal_evolution_of_energy_spectrum}
    \end{center}
\end{figure}

We can also see a similar scale-dependent behaviour in the time evolution.
Figure~\ref{fig:temporal_evolution_of_energy_spectrum} shows instantaneous spectra $E(k, t)$ at
$t /\tau_{\eta} = 0$, $20$ and $45$ corresponding to the times shown in
figure~\ref{fig:temporal_evolution_of_turbulence_attenuation}. The attenuation
of $E(k, t)$ develops from higher to lower wave-numbers over time, which is
consistent with the visualisations of the attenuation of vortices
(figure~\ref{fig:temporal_evolution_of_turbulence_attenuation}).

\begin{figure}
    \begin{center}
        \begin{overpic}[tics=10]{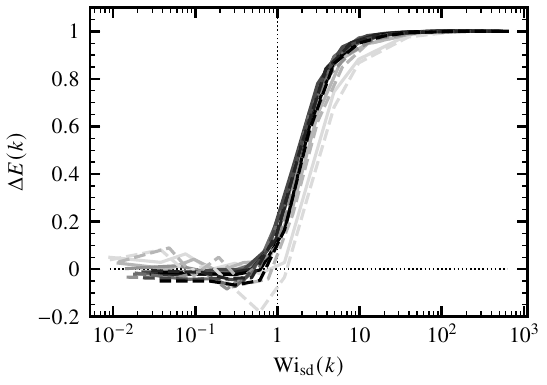}
        \end{overpic}
        \caption{
        Relative attenuation rate $\Delta E(k)$ of the energy
        spectrum. Same as in
        figure~\ref{fig:energy_attenuation_rate_afo_Wi}  but as
        a function of $\Wi_{\mathrm{sd}}(k)$ instead of $\Wi$.
        The solid and dashed lines show the results for
        $\vb*{f}_{\mathrm{ex}}^{(\mathrm{CI})}$ and
        $\vb*{f}_{\mathrm{ex}}^{(\mathrm{OU})}$, respectively.
        }
        \label{fig:energy_attenuation_rate_afo_Wi_local}
    \end{center}
\end{figure}

\subsection{Scale-dependent Weissenberg number}
\label{subsec:scale_local_weissenberg_number}
In the previous subsections, we have shown in figure~\ref{fig:vorticies_visualization} that
coherent vortices at different scales are attenuated depending on $\Wi$.
We have also shown
in figures \ref{fig:energy_spectrum} and \ref{fig:energy_attenuation_rate_afo_Wi} that
the wave-number range of the attenuated energy spectrum depends on $\Wi$.
In other words, the critical value of $\Wi$ for the turbulence attenuation depends on the scale. In
this subsection, we explain this scale-dependence in terms of the
multiple timescales of turbulence.

We estimate the turnover time of the vortices at scale corresponding to a wave-number $k$ as
\begin{equation}
    \taufk = \epsilon^{-1/3}k^{-2/3},
    \label{eq:timescale_of_turbulence}
\end{equation}
and define the scale-dependent Weissenberg number as
\begin{equation}
    \Wi_{\mathrm{sd}}(k) = \frac{\taup}{\taufk} = \frac{\taup}{\epsilon^{-1/3}k^{-2/3}}.
    \label{eq:weissenberg_number_local_afo_spectrum}
\end{equation}
When $\Wi_{\mathrm{sd}}(k) = 1$, the turnover time of
vortices at the scale $k^{-1}$ is equal to the polymer relaxation time.
In other words, the length scale $k^{-1}$ satisfying $\Wi_{\mathrm{sd}}(k) = 1$ is equivalent to the Lumley scale $\epsilon^{1/2}\taup^{3/2}$ \citep{Lumley_1969,Lumley_1973}. While the Lumley scale indicates a single threshold, $\Wi_{\mathrm{sd}}(k)$ serves as a continuous metric that quantifies the timescale ratio at each scale.
For $k = \eta^{-1}$, $\Wi_{\mathrm{sd}}(k)$ is identical to $\Wi$ defined by
\eqref{eq:weissenberg_number}. We replot, in
figure~\ref{fig:energy_attenuation_rate_afo_Wi_local}, $\Delta E(k)$ as a
function of $\Wi_{\mathrm{sd}}(k)$ instead of $\Wi$. Here, the solid and dashed
lines indicate the results for turbulence driven by
$\vb*{f}_{\mathrm{ex}}^{(\mathrm{CI})}$ and
$\vb*{f}_{\mathrm{ex}}^{(\mathrm{OU})}$, respectively. The key observation
is that, irrespective of scale and forcing, $\Delta E(k)$ gets
larger at $\Wi_{\mathrm{sd}}(k) \approx 1$. This quantitatively shows that
turbulent energy at the scale $k^{-1}$ starts to be attenuated when
the polymer relaxation time becomes comparable to the turnover time
of the $k^{-1}$-scale vortices.

Thus, $\Wi_{\mathrm{sd}}(k)$ can provide a unified description of the
onset of attenuation across different scales. This is based on the
timescale matching between polymers and multiscale vortices. We
next examine how this timescale matching is related to
polymer stretching.

\subsection{Polymer stretching}
\label{subsec:polymer_alignment}
\begin{figure}
    \begin{center}
        \begin{overpic}[tics=10]{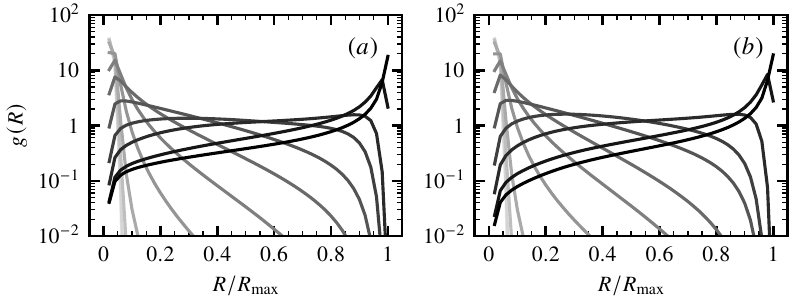}
        \end{overpic}
        \caption{
        Probability density function $g(R)$ of the polymer length $R$
        in turbulence driven by the external forces ($a$)
        $\vb*{f}_{\mathrm{ex}}^{(\mathrm{CI})}$ and ($b$)
        $\vb*{f}_{\mathrm{ex}}^{(\mathrm{OU})}$. The lines, from lighter to darker
        colour, correspond to $\Wi = 0.06$, $0.25$, $0.5$, $1$, $2$, $4$, $8$, $16$,
        $32$, $64$, $256$ and $1024$.  }
        \label{fig:polymer_length_ratio_Wi_dependence}
    \end{center}
\end{figure}
\begin{figure}
    \begin{center}
        \begin{overpic}[tics=10]{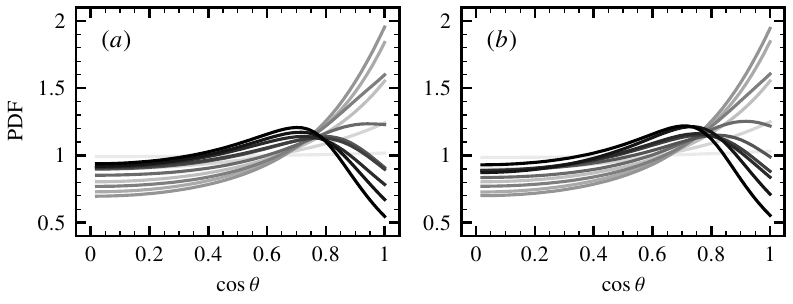}
        \end{overpic}
        \caption{
        Probability density function of the cosine $\cos
            \theta$ of the angle $\theta$ $(0 \leq \theta \leq \pi / 2)$ between the end-to-end vector $\vb*{R}$ and the strongest elongation direction $\vb*{e}_{\mathrm{max}}$
        in turbulence driven by the external forces ($a$)
        $\vb*{f}_{\mathrm{ex}}^{(\mathrm{CI})}$ and ($b$)
        $\vb*{f}_{\mathrm{ex}}^{(\mathrm{OU})}$. The lines, from
        lighter to darker colour, correspond to $\Wi = 0.06$,
        $0.25$, $0.5$, $1$, $2$, $4$, $8$, $16$, $32$, $64$,
        $256$ and $1024$.
        }
        \label{fig:polymer_alignment_no_filtered_Wi_dependence}
    \end{center}
\end{figure}

Since polymers can affect turbulence only while being sufficiently
stretched, we first confirm that they stretch significantly for $\Wi
    \gtrsim 1$. Figure~\ref{fig:polymer_length_ratio_Wi_dependence}
shows the probability density function (PDF) $g(R)$ of the polymer
length $R\:(= \abs{\vb*{R}})$ in the statistically steady state.
When $\Wi \lesssim 1$
(lighter lines), $g(R)$ has a peak around $R / R_{\mathrm{max}}
    \approx 0$ (more precisely, $R \approx
    \sqrt{3}R_{\mathrm{eq}}$), meaning that polymers remain in the coil
state. In contrast, as $\Wi$ increases, the peak shifts from $R /
    R_{\mathrm{max}} \approx 0$ to $1$, showing that polymers are stretched
for $\Wi \gtrsim 1$.

Let us examine how polymer stretching is related to turbulent
structures. We quantify this relationship using the alignment
between the end-to-end vector $\vb*{R}$ of a polymer and the
strongest stretching direction $\vb*{e}_{\mathrm{max}}$ of the
strain-rate tensor $\vb*{S}$.
Figure~\ref{fig:polymer_alignment_no_filtered_Wi_dependence} shows
the PDF of $\cos \theta\:\{=
    \abs{\vb*{e}_{\mathrm{max}}\cdot\vb*{R}} /
    (\abs{\vb*{e}_{\mathrm{max}}}\abs{\vb*{R}})\}$, where $\theta$ $(0 \leq \theta \leq \pi/2)$ is
the angle between $\vb*{e}_{\mathrm{max}}$ and $\vb*{R}$. We see
that for $\Wi \lesssim 1$ (lighter lines), the PDF is almost flat.
As $\Wi\:(\gtrsim 1)$ increases, a peak appears at $\cos \theta \approx 1$, reaching its maximum for $\Wi = 4$. For even larger $\Wi$, this alignment becomes weaker (see darker lines). These observations
mean that polymers with $\Wi \approx 1$ are most likely to align
with the stretching direction of the smallest-scale structures; note
that $\vb*{S}$ and therefore its stretching direction $\vb*{e}_{\mathrm{max}}$ is mainly determined by the smallest scale in
turbulence, because the magnitude of the velocity gradient
is larger for smaller scales. Such alignment behaviour is consistent with previous
studies \citep{Rehman_et_al_2022,Koide_and_Goto_2024}.

\begin{figure}
    \begin{center}
        \begin{overpic}[tics=10]{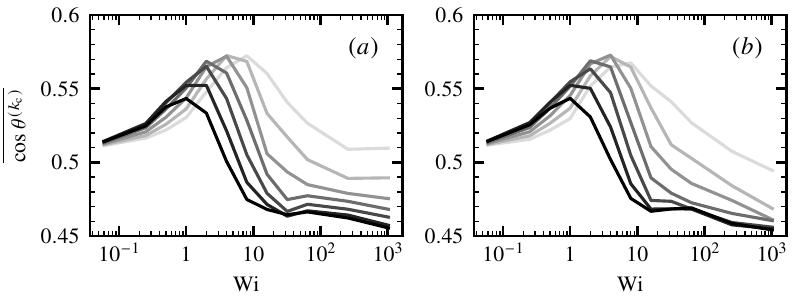}
        \end{overpic}
        \caption{
        Average $\overline{\cos \theta^{(k_{\mathrm{c}})}}$ of $\cos \theta^{(k_{\mathrm{c}})}$ as a
        function of $\Wi$ for turbulence driven by the external forces ($a$)
        $\vb*{f}_{\mathrm{ex}}^{(\mathrm{CI})}$ and ($b$)
        $\vb*{f}_{\mathrm{ex}}^{(\mathrm{OU})}$. The lines, from lighter to darker,
        correspond to the cutoff wave-numbers $k_{\mathrm{c}} \eta = 0.06$, $0.085$, $0.12$, $0.17$, $0.24$,
        $0.34$ and $0.48$.
        }
        \label{fig:expected_value_of_polymer_alignment_angle_Wi_dependence}
    \end{center}
\end{figure}

\begin{figure}
    \begin{center}
        \begin{overpic}[tics=10]{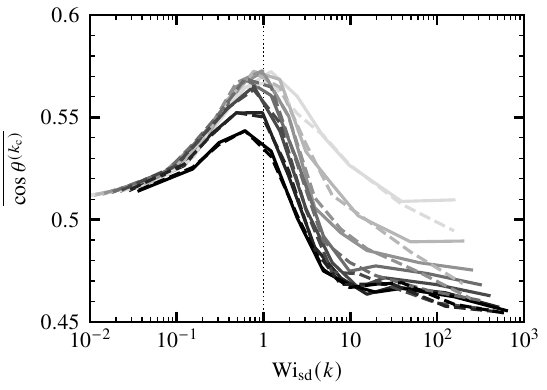}
        \end{overpic}
        \caption{
        Average $\overline{\cos \theta^{(k_{\mathrm{c}})}}$ of $\cos \theta^{(k_{\mathrm{c}})}$. Same as in
        figure~\ref{fig:expected_value_of_polymer_alignment_angle_Wi_dependence} but as a function of $\Wi_{\mathrm{sd}}(k)$. The solid and dashed lines show the
        results for $\vb*{f}_{\mathrm{ex}}^{(\mathrm{CI})}$ and
        $\vb*{f}_{\mathrm{ex}}^{(\mathrm{OU})}$, respectively.
        }
        \label{fig:expected_value_of_polymer_alignment_angle_Wi_local_dependence}
    \end{center}
\end{figure}

We next examine the alignment with the strongest stretching
directions at different scales. This examination is important because, as shown
in
\S\S\:\ref{subsec:vortices_visualization}--\ref{subsec:scale_local_weissenberg_number},
turbulence consists of multiscale vortices, and the attenuation of
these vortices is relevant to the scale-dependent attenuation of
turbulent energy. We evaluate the strain-rate
tensor $\vb*{S}^{(k_{\mathrm{c}})}$ from the band-pass filtered velocity field to
obtain the strongest stretching direction
$\vb*{e}_{\mathrm{max}}^{(k_{\mathrm{c}})}$ of $\vb*{S}^{(k_{\mathrm{c}})}$. First, we have confirmed that polymers with larger $\Wi$ tend to align with the larger-scale stretching by plotting the PDF of $\cos \theta^{(k_{\mathrm{c}})}$ of the angle between
$\vb*{R}$ and $\vb*{e}_{\mathrm{max}}^{(k_{\mathrm{c}})}$ (figures are omitted). To demonstrate this tendency, we show in
figure~\ref{fig:expected_value_of_polymer_alignment_angle_Wi_dependence}
the average $\overline{\cos \theta^{(k_{\mathrm{c}})}}$ of
$\cos \theta^{(k_{\mathrm{c}})}$ over polymers and time as a function of
$\Wi$. Values above $0.5$ indicate preferential alignment with the
$k^{-1}$-scale stretching direction. The lines, from lighter to
darker, correspond to the different cutoff wave-numbers
$k_{\mathrm{c}}$ of the filter: $k_{\mathrm{c}} \eta = 0.06$,
$0.085$, $0.12$, $0.17$, $0.24$, $0.34$ and $0.48$.
We can observe that the darkest line has a peak at $\Wi
    \approx 1$ while the lighter lines show peaks at the larger $\Wi$.
This means that polymers with longer $\taup$ (i.e.~larger $\Wi$)
align with larger-scale stretching directions. To show that
this tendency can be also explained in terms of timescale matching
between polymers and turbulence, we replot, in
figure~\ref{fig:expected_value_of_polymer_alignment_angle_Wi_local_dependence},
$\overline{\cos \theta^{(k_{\mathrm{c}})}}$ as a function of
$\Wi_{\mathrm{sd}}(k)$ instead of $\Wi$. The solid and dashed lines
are the results for the turbulence driven by
$\vb*{f}_{\mathrm{ex}}^{(\mathrm{CI})}$ and
$\vb*{f}_{\mathrm{ex}}^{(\mathrm{OU})}$, respectively. We see that
$\overline{\cos \theta^{(k_{\mathrm{c}})}}$
show peaks at $\Wi_{\mathrm{sd}}(k) \approx 1$ irrespective of the
scale and forcing. Thus, polymers align with the timescale-matched
vortices. This supports the physical picture that turbulence
attenuation occurs when polymers are stretched and align with vortices
whose turnover time is comparable to the polymer relaxation time.

\section{Discussion}
\label{sec:Discussion}
Considering these observations and statistical results presented in \S\:\ref{sec:Turbulence_attenuation} collectively, we detail the timescale-based physical mechanism of turbulence attenuation (\S\:\ref{subsec:Physical_mechanism}). This mechanism is clearly revealed only through $\Wi_{\mathrm{sd}}(k)$. Furthermore, building on this timescale perspective, we link the spectral behaviour to the scale-by-scale interaction between polymers and multiscale vortices (\S\:\ref{subsec:spectral_behaviour}).

\subsection{Scale-by-scale mechanism of turbulence attenuation}
\label{subsec:Physical_mechanism}
We summarise the physical mechanism of turbulence attenuation by
polymers in terms of the scale-dependent Weissenberg number $\Wi_{\mathrm{sd}}(k)$.
First, we can describe the temporal evolution of the transient attenuation process (figure~\ref{fig:temporal_evolution_of_turbulence_attenuation}).
Specifically, polymers with $\taup (\gtrsim \tau_{\eta})$ are initially stretched by the smallest-scale vortices, because these vortices stretch polymers fastest. Consequently, these smallest-scale vortices are suppressed [figure~\ref{fig:temporal_evolution_of_turbulence_attenuation}($b$)]. Subsequently, the polymers suppress vortices with the shortest turnover time among the remaining coherent vortices [figure~\ref{fig:temporal_evolution_of_turbulence_attenuation}($c$)], extracting turbulent energy from these vortices (figure~\ref{fig:temporal_evolution_of_energy_spectrum}). This scale-by-scale interaction between polymers and multiscale vortices therefore proceeds successively from smaller to larger scales. This process ends at scales corresponding to $\taup \approx \tau_{\mathrm{f}}(k)$ (i.e.~$\Wi_{\mathrm{sd}}(k) \approx 1$), since polymers relax more rapidly than they are stretched by vortices at scales where $\Wi_{\mathrm{sd}}(k) \lesssim 1$.

Thus, in the statistically steady state,
turbulence is attenuated at scales satisfying $\Wi_{\mathrm{sd}}(k) \gtrsim 1$ (figure~\ref{fig:vorticies_visualization}). Since smaller-scale vortices are already suppressed, polymers interact most effectively with the remaining vortices whose turnover time is comparable to the polymer relaxation time (i.e.~$\Wi_{\mathrm{sd}}(k) \approx 1$). As a result, turbulent energy contained in these vortices begins to attenuate (figures~\ref{fig:energy_spectrum} and \ref{fig:energy_attenuation_rate_afo_Wi}). This occurs because polymers align with the stretching direction induced by the timescale-matched vortices (figures~\ref{fig:expected_value_of_polymer_alignment_angle_Wi_dependence} and \ref{fig:expected_value_of_polymer_alignment_angle_Wi_local_dependence}), extracting turbulent energy from them. Therefore, the scale-dependent Weissenberg number $\Wi_{\mathrm{sd}}(k)$ successfully explains the onset of scale-dependent turbulence attenuation.

So far, we have considered the case with large $\xi$ and
$L_{\mathrm{max}}$, where polymers can store sufficient elastic
energy (appendix~\ref{sec:APPENDIX_energy}). Here, it
is worth emphasising that $\Wi_{\mathrm{sd}}(k) \approx 1$
remains an important indicator for the onset of turbulence
attenuation even when polymers store less elastic energy, i.e.~for
smaller $\xi$ and $L_{\mathrm{max}}$. We show results for these
cases in
appendix~\ref{sec:APPENDIX_Dependence_on_the_viscosity_ratio_and_maximum_length_of_polymer}.
These results show that for smaller $\xi$ and $L_{\mathrm{max}}$,
the magnitude of turbulence attenuation does become smaller, but the
scale-dependent criterion for the turbulence attenuation remains
governed by $\Wi_{\mathrm{sd}}(k)$.

\subsection{Physical understanding of the spectral behaviour}
\label{subsec:spectral_behaviour}
\begin{figure}
    \begin{center}
        \begin{overpic}[tics=10]{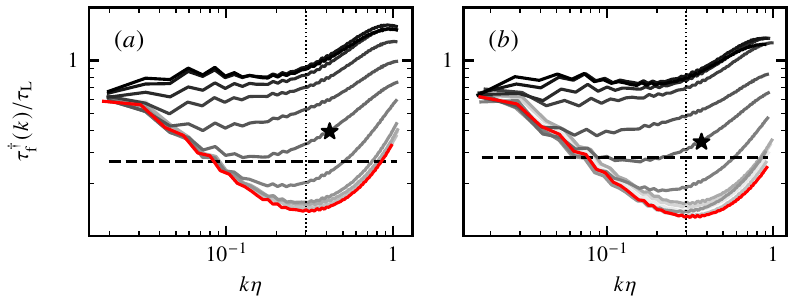}
        \end{overpic}
        \caption{
        Turnover time $\tau^{\dagger}_{\mathrm{f}}(k) / \tauL$ of vortices in the polymer solution turbulence as a function of wave-number $k\eta$, driven by the external forces ($a$) $\vb*{f}_{\mathrm{ex}}^{(\mathrm{CI})}$ and ($b$) $\vb*{f}_{\mathrm{ex}}^{(\mathrm{OU})}$. The lines, from lightest to darkest, correspond to Weissenberg numbers $\Wi = 0.06$, $0.25$, $0.5$, $1$, $2$, $4$, $8$, $16$, $32$, $64$, $256$ and $1024$. $\tauL$ is the integral time shown in table~\ref{tab:dns_parameters}. The red line represents $\tau^{\dagger}_{\mathrm{f}}(k)$ for Newtonian turbulence. The line marked with a star represents $\tau^{\dagger}_{\mathrm{f}}(k)$ for $\Wi = 8$, and the horizontal dashed line indicates the corresponding value of $0.5\taup / \tauL$.
        }
        \label{fig:timescale_of_turbulence_afo_k}
    \end{center}
\end{figure}

\begin{figure}
    \begin{center}
        \begin{overpic}[tics=10, width=1.7in]{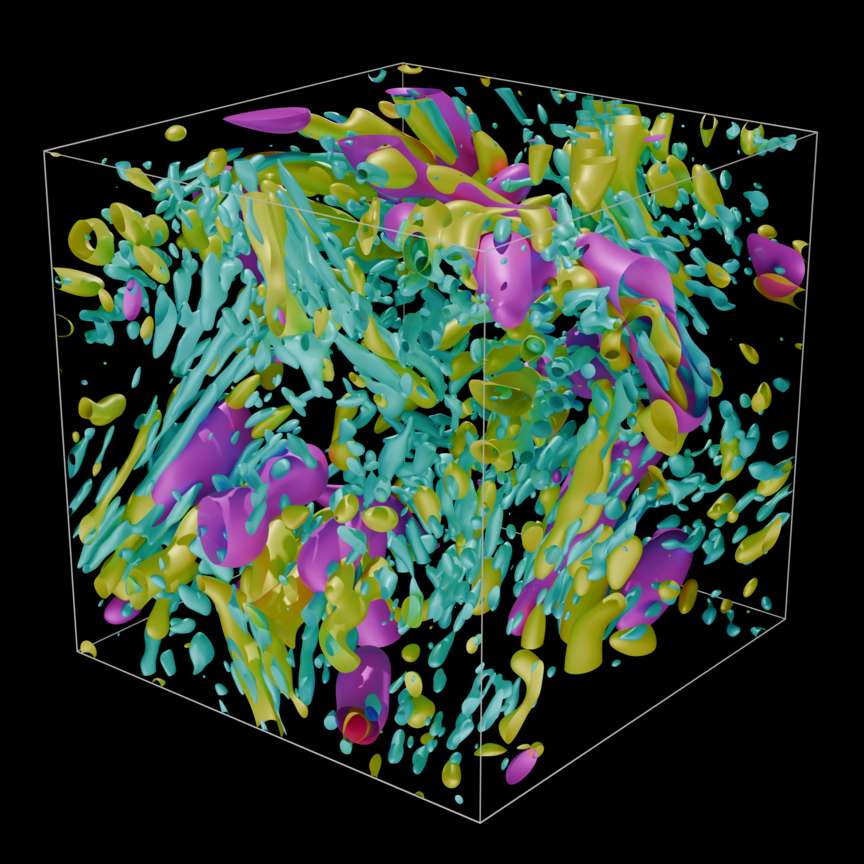}
            \put(4.25, 15.5){\includegraphics[width=1cm]{axis.pdf}}
        \end{overpic}
        \caption{
            Positive isosurfaces of the second invariant $Q^{(k_{\mathrm{c}})}$ of the velocity gradient
            tensor at three different scales $k_{\mathrm{c}}\eta = 0.06$ (magenta),
            $0.12$ (yellow) and $0.24$ (cyan) for the polymer solution at $\Wi = 1024$ for a statistically steady state. The threshold at each scale is $Q^{(k_{\mathrm{c}})} = 0.14$ (twice the standard deviation of $Q^{(k_{\mathrm{c}})}$ at $k_{\mathrm{c}}\eta = 0.24$ for $\Wi = 1024$).
        }
        \label{fig:vortices_visualisation_same_threshold}
    \end{center}
    \begin{center}
        \begin{overpic}[tics=10]{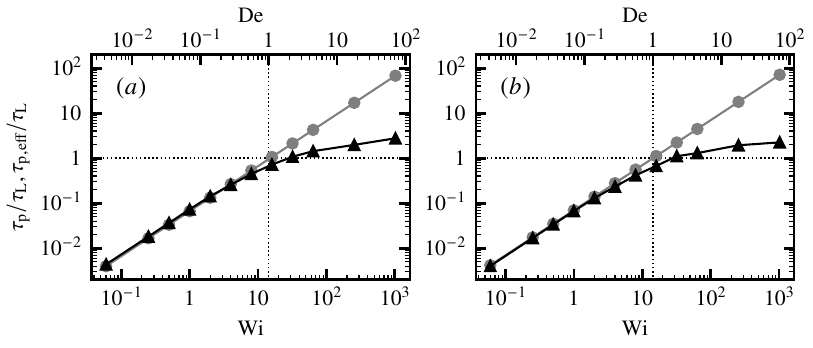}
        \end{overpic}
        \caption{
        Polymer relaxation time $\taup$ (grey circles) and the length-dependent relaxation time $\tau_{\mathrm{p, eff}}(R)$ (black triangles) as functions of $\Wi$ and $\De$ in turbulence driven by the external forces ($a$) $\vb*{f}_{\mathrm{ex}}^{(\mathrm{CI})}$ and ($b$) $\vb*{f}_{\mathrm{ex}}^{(\mathrm{OU})}$.
        }
        \label{fig:polymer_relaxation_time_compare}
    \end{center}
\end{figure}

Next, to link the scale-by-scale attenuation process to the spectral behaviour (\S\:\ref{subsec:energy_spectrum}), we investigate the characteristic timescales of attenuated vortices at different scales in turbulent polymer solutions. Here, we evaluate the turnover time $\tau^{\dagger}_{\mathrm{f}}(k)$ of the $k^{-1}$-scale vortices in turbulence attenuated by polymers as
\begin{gather}
    \tau^{\dagger}_{\mathrm{f}}(k) = \qty{k^3E(k)}^{-1/2}. \label{eq:timescale_of_turbulence_based_on_spectra}
\end{gather}
Note that although for the Newtonian turbulence (i.e.~$E(k) \propto k^{-5/3}$), \eqref{eq:timescale_of_turbulence_based_on_spectra} yields the same scaling as \eqref{eq:timescale_of_turbulence}, it expresses the realistic timescale of $k^{-1}$-scale vortices in the attenuated turbulence. In figure~\ref{fig:timescale_of_turbulence_afo_k}, we plot $\tau^{\dagger}_{\mathrm{f}}(k)$ as a function of $k$ for turbulence driven by ($a$) $\vb*{f}_{\mathrm{ex}}^{(\mathrm{CI})}$ and ($b$) $\vb*{f}_{\mathrm{ex}}^{(\mathrm{OU})}$ for different values of $\Wi$. When $\Wi \ll 1$ (lightest line), we observe $\tau^{\dagger}_{\mathrm{f}}(k) \propto k^{-2/3}$ for the wave-numbers $k\eta \lesssim 0.3$, indicating that smaller vortices possess shorter turnover times. In contrast, as $\Wi$ gets larger than unity (darker lines), $\tau^{\dagger}_{\mathrm{f}}(k)$ deviates from the red line for the Newtonian turbulence in a higher wave-number range. It is important to observe that $\tauf^{\dagger}(k)$ exhibits a plateau in a lower wave-number range, and that its level is approximately $\taup$ when $\tau_{\eta} < \taup < \tauL$ (e.g. for $\Wi = 8$; see the line marked with a star). These behaviours of $\tau^{\dagger}_{\mathrm{f}}(k)$, i.e. the plateaus and their levels, can be understood by the picture that polymers suppress vortices satisfying $\Wi_{\mathrm{sd}}(k) \gtrsim 1$. To this end, we define another scale-dependent Weissenberg number as $\Wi^{\dagger}_{\mathrm{sd}}(k) = \taup / \tau^{\dagger}_{\mathrm{f}}(k)$ by using $\tau^{\dagger}_{\mathrm{f}}(k)$. For wave-numbers $k$ satisfying $\Wisdk \gtrsim 1$, polymers attenuate turbulent energy spectrum $E(k)$, which corresponds to an increase in $\tau^{\dagger}_{\mathrm{f}}(k)$ and therefore decrease in $\Wi^{\dagger}_{\mathrm{sd}}(k)$. This process continues until $\Wi^{\dagger}_{\mathrm{sd}}(k)$ gets as small as unity. Thus, $\tau^{\dagger}_{\mathrm{f}}(k)$ increases as large as $\taup$ independent of the wave-number $k$, which implies the plateaus of $\tau^{\dagger}_{\mathrm{f}}(k)$ and the $k^{-3}$ scaling of $E(k)$.

Incidentally, it is interesting to observe in figure~\ref{fig:vortices_visualisation_same_threshold} that the multiscale vortex structures of attenuated turbulence are still observed for the same threshold of $Q^{(k_{\mathrm{c}})}$ at each scale. However, the coherence of these structures are qualitatively different from those in the Newtonian case. Small- and medium-scale vortices seem to align in the same direction. This implies that the nonlinear interactions representative of vortex stretching are suppressed. Thus, this turbulent state for $\De > 1$ might be understood as analogous to the three-dimensional Newtonian turbulence without vortex stretching investigated by \citet{Bos_2021}. In fact, it has also been reported that turbulence without vortex stretching exhibits drag reduction similar to that observed in polymer solutions \citep{Bos_2025}.

As shown in figure~\ref{fig:timescale_of_turbulence_afo_k} (darker lines), $\tau^{\dagger}_{\mathrm{f}}(k)$ does not increase beyond $\mathcal{O}(\tau_{\mathrm{L}})$, even though $\taup \gtrsim \tauL$ (i.e.~$\De \gtrsim 1$). This is because there are no vortices with turnover times longer than $\tauL$. Consequently, one might think that the physical picture based on $\Wi^{\dagger}_{\mathrm{sd}}(k)$ becomes invalid in this regime. To demonstrate that this physical picture is valid for $\De > 1$, we need to consider the effect of nonlinear elongation on polymer relaxation dynamics through an effective relaxation timescale $\tau_{\mathrm{p, eff}}(R)$. Here, we define this effective relaxation time $\tau_{\mathrm{p, eff}}(R)$ as
\begin{gather}
    \tau_{\mathrm{p, eff}}(R) = \tau_{\mathrm{p}} \frac{\left\langle R^2 \right\rangle}{ \left\langle R^2 / (1 - R^2 / R_{\mathrm{max}}^2) \right\rangle}. \label{eq:effective_polymer_relaxation_time}
\end{gather}
The rationale for this definition is provided in appendix~\ref{sec:APPENDIX_length_dependent_relaxation_time}. The importance of the effective relaxation time, which varies with polymer elongation, has been demonstrated under steady extensional flow \citep{Watanabe_and_Matsumiya_2017}. In figure~\ref{fig:polymer_relaxation_time_compare}, we show $\taup$ and $\tau_{\mathrm{p, eff}}(R)$ normalized by $\tauL$ as a function of $\Wi$ or $\De$. For $\De \lesssim 1$, $\tau_{\mathrm{p, eff}}(R)$ is approximately equal to $\taup$. In stark contrast, for $\De \gtrsim 1$, $\tau_{\mathrm{p, eff}}(R)$ is shorter than $\taup$. Furthermore, for $\De \gg 1$, $\tau_{\mathrm{p, eff}}(R)$ reaches a constant value of $\mathcal{O}(\tauL)$ (dotted line). By defining the effective scale-dependent Weissenberg number as $\Wi^{\dagger}_{\mathrm{sd, eff}}(k) = \tau_{\mathrm{p, eff}}(R) / \tau^{\dagger}_{\mathrm{f}}(k)$, we find that the criterion $\Wi^{\dagger}_{\mathrm{sd, eff}}(k) \approx \mathcal{O}(1)$ is satisfied. Thus, we conclude that the physical picture based on $\Wi^{\dagger}_{\mathrm{sd}}(k)$ is valid across a wide range of $\Wi$.

\subsection{Spectral scaling law}
\label{subsec:Discrepancy}
In this subsection, we discuss the spectral scaling law such as the $k^{-3}$ scaling and the different scaling reported in recent studies. Some of the previous experimental and numerical results have suggested the $k^{-3}$ scaling, which is observed in the present simulations at $\Re_{\lambda} \approx 150$. For instance, \citet{Vonlanthen_and_Monkewitz_2013} observed this $k^{-3}$ scaling in their experimental investigation of grid turbulence in dilute polymer solutions at $\Re_{\lambda} \approx 100$ and different concentrations. Similarly, \citet{Valente_et_al_2014, Valente_et_al_2016} reported the $k^{-3}$ spectrum in their direct numerical simulations of homogeneous isotropic turbulence using the FENE-P model at $\Re_{\lambda} \approx 50$--$400$ and $\Wi \approx 0.5$--$80$. In contrast, recent studies have observed different scaling laws. For example, \citet{Zhang_et_al_2021} experimentally measured von K\'{a}rm\'{a}n turbulence at $\Re_{\lambda} = 350$--$530$ and showed that the second-order velocity structure function follows $r^{1.38}$, corresponding to $E(k) \propto k^{-2.38}$. Likewise, \citet{Rosti_et_al_2023} and \citet{Chiarini_et_al_2026} conducted direct numerical simulations at $\Re_{\lambda} \approx 400$ using Oldroyd-B and FENE-P models and demonstrated $E(k) \propto k^{-2.3}$. Several factors may contribute to the difference in scaling exponents, including the Reynolds number, Weissenberg number, polymer--polymer interaction, hydrodynamic interaction, degradation and polymer models. More importantly, rather than merely focusing on the scaling exponent of the spectrum, it is essential to understand the underlying physics. Since the present framework is applicable to any turbulent flow and polymer model, further studies are required to evaluate how the above factors alter the physical mechanism of polymer--turbulence interaction from a timescale perspective.


\section{Conclusion}
\label{sec:CONCLUSION}
In the present study, we have investigated the physical mechanism of turbulence attenuation by polymers from a timescale perspective. To this end, we have conducted DNS of homogeneous isotropic turbulence in polymer solutions using an Eulerian--Lagrangian approach. We model polymers as FENE dumbbells and track their motions using Brownian dynamics simulations. The polymer--turbulence interaction is represented by the force exerted on the fluid by the polymer elastic stress. We make a fundamental contribution by providing a novel timescale-based framework using scale decomposition that successfully captures the physical picture underlying turbulence attenuation by polymers.

In \S~\ref{sec:Turbulence_attenuation}, we have discussed turbulence attenuation. Polymers with a relaxation time longer than the Kolmogorov time suppress small-scale vortices (figure~\ref{fig:vorticies_visualization}) and attenuate the turbulent energy contained by these vortices (figures~\ref{fig:energy_spectrum} and \ref{fig:energy_attenuation_rate_afo_Wi}). To clarify the physical mechanism of turbulence attenuation, we define the scale-dependent Weissenberg number $\Wi_{\mathrm{sd}}(k)$ \eqref{eq:weissenberg_number_local_afo_spectrum} as the ratio of the polymer relaxation time to the vortex turnover time. In the statistically steady state, the attenuated scales satisfy the condition $\Wi_{\mathrm{sd}}(k) \gtrsim 1$ (figure~\ref{fig:energy_attenuation_rate_afo_Wi_local}). In physical space, polymers are more likely to align with the stretching direction induced by vortices at scales satisfying $\Wi_{\mathrm{sd}}(k) \approx 1$ (figures~\ref{fig:expected_value_of_polymer_alignment_angle_Wi_dependence} and \ref{fig:expected_value_of_polymer_alignment_angle_Wi_local_dependence}). We further investigated the transient attenuation process, and demonstrate that polymers initially suppress the smallest-scale vortices and subsequently suppress larger-scale ones (figures~\ref{fig:temporal_evolution_of_turbulence_attenuation} and \ref{fig:temporal_evolution_of_energy_spectrum}).

In \S~\ref{subsec:Physical_mechanism}, we have described the physical mechanism of turbulence attenuation in terms of timescale matching between polymers and multiscale vortices. During the transient process, polymers with $\taup \gtrsim \tau_{\eta}$ are first stretched by the smallest-scale vortices, which can stretch the polymers fastest. As a result, these vortices are suppressed first. Subsequently, polymers suppress vortices with the shortest timescale among the remaining coherent vortices by extracting turbulent energy from them. This turbulence attenuation therefore proceeds from smaller to larger scales, and continues up to scales corresponding to $\Wi_{\mathrm{sd}}(k) \approx 1$. In \S\:\ref{subsec:spectral_behaviour}, to link the present physical mechanism to the spectral behaviour, we have shown that the realistic turnover times $\tau^{\dagger}_{\mathrm{f}}(k)$ \eqref{eq:timescale_of_turbulence_based_on_spectra} at different scales in turbulence attenuated by polymers increase with $\Wi$ (figure~\ref{fig:timescale_of_turbulence_afo_k}). Moreover, this increased $\tau^{\dagger}_{\mathrm{f}}(k)$ tends to plateau. This behaviour of $\tau^{\dagger}_{\mathrm{f}}(k)$ can be understood by defining the scale-dependent Weissenberg number as $\Wi^{\dagger}_{\mathrm{sd}}(k) = \taup / \tau^{\dagger}_{\mathrm{f}}(k)$. The key is that the attenuation process at the scale where $\Wi_{\mathrm{sd}}(k) \gtrsim 1$ continues until $\Wi^{\dagger}_{\mathrm{sd}}(k)$ gets as small as unity. Hence, $\tau^{\dagger}_{\mathrm{f}}(k)$ becomes as large as $\taup$, independent of the scale, which corresponds to $E(k) \propto k^{-3}$. Even though $\tau^{\dagger}_{\mathrm{f}}(k)$ does not increase beyond the integral timescale $\tauL$ for $\taup > \tauL$ (i.e.~$\De > 1$), by defining the effective relaxation time $\tau_{\mathrm{p, eff}}(R)$ as \eqref{eq:effective_polymer_relaxation_time}, we have clarified that the relaxation time of polymers effectively becomes shorter than $\taup$ at $\De > 1$ (figure~\ref{fig:polymer_relaxation_time_compare}). This implies that the physical picture based on $\Wi^{\dagger}_{\mathrm{sd}}(k)$ remains valid for $\De > 1$. In \S\:\ref{subsec:Discrepancy}, we have discussed the scaling law of the attenuated energy spectrum. Some previous findings are consistent with the present study, whereas others show differences that may arise from several factors. Nevertheless, the present analysis framework based on $\Wi_{\mathrm{sd}}(k)$ provides a guiding principle in terms of timescale for understanding turbulence attenuation by polymers, which will help reveal the effects of these factors in future work.

Overall, we conclude that the timescale matching between polymers and multiscale vortices plays an important role in turbulence attenuation. For a sufficiently high polymer concentration $\xi$ and large maximum length $\Lmax$ (i.e.~when polymers can store sufficient elastic energy), we can explain the physical mechanism of turbulence attenuation using the scale-dependent Weissenberg number $\Wi_{\mathrm{sd}}(k)$. Furthermore, for smaller $\xi$ and shorter $\Lmax$, although the magnitude of the attenuation decreases, the mechanism remains well described by $\Wi_{\mathrm{sd}}(k)$ (Appendix~\ref{sec:APPENDIX_Dependence_on_the_viscosity_ratio_and_maximum_length_of_polymer}).


\section*{Acknowledgements}
This study was supported by the JSPS Grant-in-Aid for JSPS Fellows (JP24KJ1627), Scientific Research (25K01158 and 26K17306) and Transformative Research Areas (JP26H00389). The numerical simulations were conducted using computational resources of Fugaku provided by the RIKEN Center for Computational Science (hp260139) and Plasma Simulatior ``Sousei'' provided under the auspices of the NIFS Collaboration Research Program (NIFS26KISC034).

\section*{Declaration of interests}
The authors report no conflict of interest.

\appendix

\section{Dependence on the number of simulated polymers}
\label{sec:APPENDIX_dependence_of_the_number_of_polymers}
\begin{figure}
    \begin{center}
        \includegraphics[width = 0.5\linewidth]{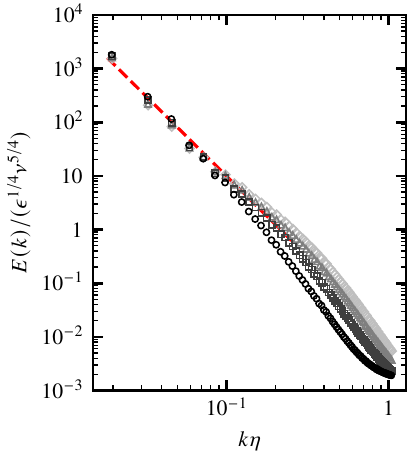}
        \caption{
        Averaged energy spectra $E(k)$ for polymer solutions in the statistically steady state of turbulence driven by $\vb*{f}^{(\mathrm{CI})}_{\mathrm{ex}}$. Circles, squares, triangles, and diamonds correspond to $\Npdash / N_{\mathrm{grid}} = 5, 10, 20$, and $40$, respectively (where $N_{\mathrm{grid}} = 512^3$ is the number of grid points). The red dashed line indicates $E(k) \propto k^{-3}$.
        }
        \label{fig:APPENDIX_energy_spectrum_dependent_of_Npdash}
    \end{center}
\end{figure}

We investigate the dependence of the number $\Npdash$ of the simulated FENE dumbbells on the spectral behaviour. Figure~\ref{fig:APPENDIX_energy_spectrum_dependent_of_Npdash} shows the energy spectrum $E(k)$ with different $\Npdash$ for polymer solutions. The parameters are $\Wi = 1024$, $\xi = 0.1$ and $\Lmax = 100$, where turbulence is significantly attenuated. The circle, square, triangle and diamond markers indicate $\Npdash / N_{\mathrm{grid}} = 5$, $10$, $20$ and $40$, respectively, where $N_{\mathrm{grid}} = 512^3$ is the number of grid point. As $\Npdash$ decreases, the magnitude of $E(k)$ decreases for the viscous dissipation range, whereas the spectrum shows $E(k) \propto k^{-3}$ outside this range irrespective of $\Npdash$ (see the red dash line). Thus, we discuss turbulence attenuation in this wave-number range with $\Npdash / N_{\mathrm{grid}} = 10$.

\section{Kinetic energy and elastic energy}
\label{sec:APPENDIX_energy}
\begin{figure}
    \begin{center}
        \includegraphics[width = \linewidth]{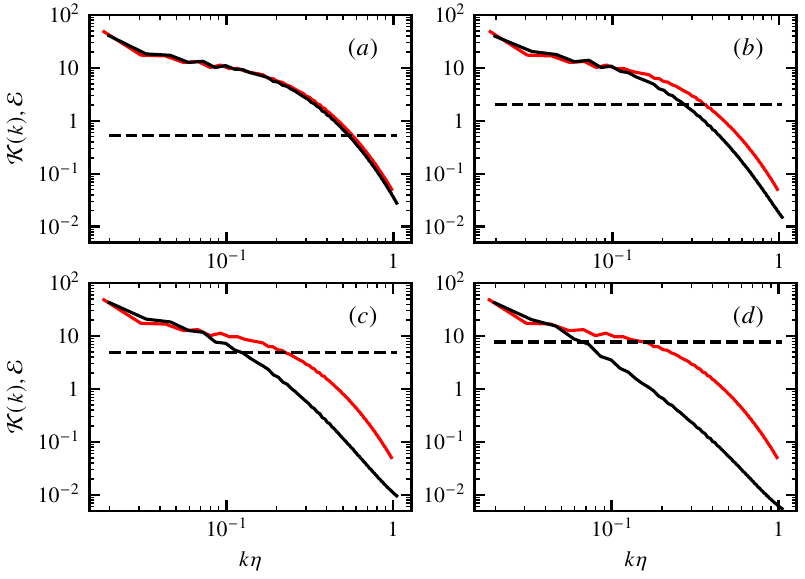}
        \caption{
            Elastic energy $\mathcal{E}$ of polymers (the dashed line) and scale-dependent turbulent energy $\mathcal{K}(k)$ of vortices at $k^{-1}$ scale (the solid
            lines) for ($a$) $\Wi = 2$, ($b$) $4$, ($c$) $8$ and ($d$) $16$. The red solid
            line shows $\mathcal{K}(k)$ for the Newtonian turbulence. The vertical axis is
            normalised by $\epsilon^{1/2}\nu^{1/2}$.
        }
        \label{fig:kinetic_energy_and_elastic_energy}
    \end{center}
\end{figure}
We show that the polymers used in the present study can store
elastic energy comparable to the turbulent energy. Using the
magnitude of elastic force
\begin{gather}
    F_{\mathrm{e}}(R) = - 2h\frac{R}{1 - R^2 / \Rmax^2}
\end{gather}
exerted on a FENE dumbbell of length $R$, we estimate the elastic energy
$U_{\mathrm{single}}(R)$ stored by a single polymer as
\begin{gather}
    U_{\mathrm{single}}(R) = - \int_{0}^{R} F_{\mathrm{e}}(R') dR' = - h \Rmax^2 \ln\qty{1 - \qty(\frac{R}{\Rmax})^2}.
\end{gather}
Therefore, the total elastic energy per unit volume in
solution is
\begin{gather}
    \mathcal{E} = \frac{N_{\mathrm{p}}}{V}\int_{0}^{\Rmax} U_{\mathrm{single}}(R)g(R) dR.
\end{gather}
Figure~\ref{fig:kinetic_energy_and_elastic_energy} shows $\mathcal{E}$ for ($a$)
$\Wi = 2$, ($b$) $4$, ($c$) $8$ and ($d$) $16$ by the horizontal black dashed
line. The black solid curve shows the scale-dependent turbulent energy
$\mathcal{K}(k)$, where $\mathcal{K}(k)\:\{=kE(k)\}$ represents the turbulent
energy contained by $k^{-1}$-scale vortices. The red solid curve shows
$\mathcal{K}(k)$ for the Newtonian turbulence. We see that for all values of $\Wi$,
$\mathcal{E}$ is comparable to $\mathcal{K}(k)$ at the wave-numbers where
turbulent energy is attenuated. This suggests that the $k^{-1}$-scale turbulent
energy is transferred to polymers and stored as elastic energy.

\section{Dependence on the viscosity ratio $\xi$ and maximum length ratio $L_{\mathrm{max}}$}
\label{sec:APPENDIX_Dependence_on_the_viscosity_ratio_and_maximum_length_of_polymer}
\begin{figure}
    \begin{center}
        \includegraphics[width = \linewidth]{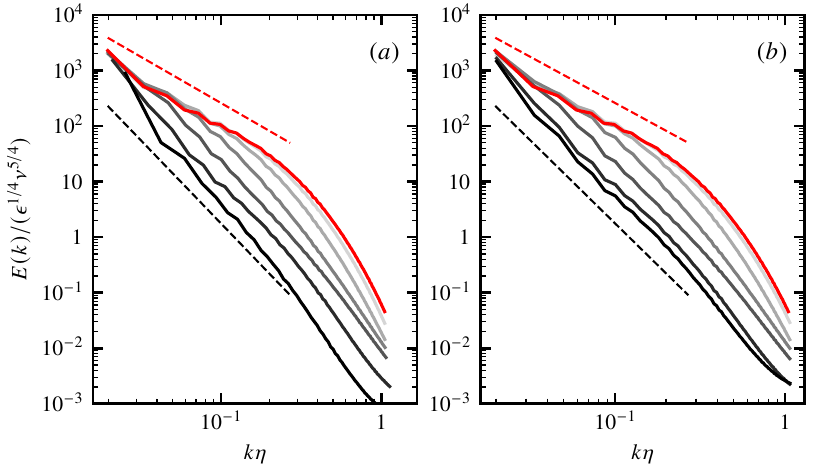}
        \caption{
        Energy spectra for ($a$) $\xi = 3.9 \times 10^{-4}$, $1.6 \times 10^{-3}$, $6.3 \times 10^{-3}$, $0.025$, $0.1$ and $0.4$ ($\Wi = 1024$, $\Lmax = 100$) and ($b$) $\Lmax = 6$, $12$, $25$, $50$, $100$ and $200$ ($\Wi = 1024$, $\xi = 0.1$) in turbulence driven by $\vb*{f}_{\mathrm{ex}}^{(\mathrm{CI})}$ ($\Re_{\lambda} \approx 150$). Darker lines correspond to larger values of $\xi$ or $\Lmax$.
        }
        \label{fig:APPENDIX_energy_spectrum_xi_and_L_dependence}
    \end{center}
\end{figure}
\begin{figure}
    \begin{center}
        \includegraphics[width = \linewidth]{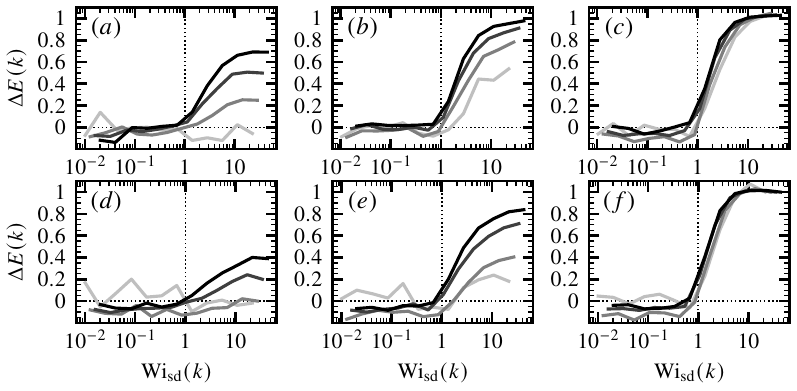}
        \caption{
            Energy attenuation rate $\Delta E(k)$ as a function of $\Wi_{\mathrm{sd}}(k)$ for ($a$--$c$) $\xi = 0.003, 0.012$ and $0.05$ ($\Lmax = 100$) and ($d$--$f$) $\Lmax = 12, 25$ and $100$ ($\xi = 0.1$) at $\Re_{\lambda} \approx 85$ and $\Wi = 128$. Darker lines correspond to higher wave-numbers $k / (\epsilon^{1/4}\nu^{-3/4}) \approx 0.08, 0.11, 0.16$ and $0.23$.
        }
        \label{fig:APPENDIX_energy_attenuation_rate_xi_and_L_dependence}
    \end{center}
\end{figure}
We discuss the effect of the viscosity ratio $\xi$ and maximum length ratio $\Lmax$ on turbulence attenuation. We show in figure~\ref{fig:APPENDIX_energy_spectrum_xi_and_L_dependence} the energy spectra with $\xi = 3.9 \times 10^{-4}$, $1.6 \times 10^{-3}$, $6.3 \times 10^{-3}$, $0.025$, $0.1$ and $0.4$ (under the fixed $\Lmax = 100$) [figure~\ref{fig:APPENDIX_energy_spectrum_xi_and_L_dependence}($a$)] and with $\Lmax = 6$, $12$, $25$, $50$, $100$ and $200$ (under the fixed $\xi = 0.1$) [figure~\ref{fig:APPENDIX_energy_spectrum_xi_and_L_dependence}($b$)]. The darker lines indicate larger $\xi$ or $\Lmax$. The Weissenberg number is $\Wi = 1024$ and Taylor microscale Reynolds number is $\Re_{\lambda} \approx 150$. Since $\Wi = 1024$ corresponds to $\De \approx 68.2$, the relaxation time of polymers is sufficiently longer than the turnover time of eddies at all scales; that is, $\Wi_{\mathrm{sd}}(k) \gtrsim 1$ is satisfied. We thus see in the figure that for large values of $\xi$ or $\Lmax$, $E(k)$ is attenuated more significantly over the wide range of wave-numbers (darker lines). In contrast, for smaller $\xi$ or $\Lmax$, the attenuation is less pronounced (lighter lines).

However, we can see that even for these smaller values of $\xi$ and $\Lmax$, turbulence attenuation occurs at the scales where $\Wi_{\mathrm{sd}}(k) \gtrsim 1$. Figure~\ref{fig:APPENDIX_energy_attenuation_rate_xi_and_L_dependence} shows the energy attenuation rate $\Delta E(k)$ as a function of $\Wi_{\mathrm{sd}}(k)$ for turbulence driven by $\vb*{f}_{\mathrm{ex}}^{\mathrm{(CI)}}$ with values of $\xi = $ ($a$)\ $0.003$, ($b$)\ $0.012$ and ($c$)\ $0.05$, and of $\Lmax = $ ($d$)\ $12$,  ($e$)\ $25$ and ($f$)\ $100$. Although for smaller $\xi$ and $\Lmax$, the magnitude of the energy attenuation rate is lower than that for larger $\xi$ and $\Lmax$, it tends to increase where $\Wi_{\mathrm{sd}}(k) \gtrsim 1$. This suggests that the timescale-matching condition holds independently of $\xi$ and $\Lmax$.

\section{Effective relaxation time of polymers}
\label{sec:APPENDIX_length_dependent_relaxation_time}
The formulation \eqref{eq:effective_polymer_relaxation_time} of the effective relaxation time $\tau_{\mathrm{p, eff}}(R)$ is derived from the balance between stretching by fluid and relaxation by polymers. Specifically, we take the inner product of $\vb*{R}$ with the governing equation \eqref{eq:langevin_equation_of_R_vector} for the end-to-end vector $\vb*{R}$ and average over space and time. We then obtain the time evolution equation for the mean square of $\vb*{R}$ as
\begin{align}
    \dv{\left\langle R^2 \right\rangle}{t} & = 2\left\langle\qty(\grad{\vb*{u}}(\vb*{r}_{\mathrm{g}}) \cdot \vb*{R}) \cdot \vb*{R}\right\rangle - \left\langle \frac{1}{\tau_{\mathrm{p}}} \frac{R^2}{1 - R^2 / R_{\mathrm{max}}^2} \right\rangle + \frac{3R_{\mathrm{eq}}^2}{\taup} \notag \\
                                           & = \Gamma_{\mathrm{fluid}} +  \Gamma_{\mathrm{elast}} + \Gamma_{\mathrm{thermo}}.
    \label{eq:avaraged_langevin_equation_of_R_vector}
\end{align}
Here, $\Gamma_{\mathrm{fluid}}$, $\Gamma_{\mathrm{elast}}$ and $\Gamma_{\mathrm{thermo}}$ represent the contributions of the fluid, elastic and thermal fluctuation forces, respectively. A positive (negative) value of each term contributes to stretching (relaxation). The elastic term $\Gamma_{\mathrm{elast}}$ contributes only to relaxation, since its value is negative. In the steady state, the left-hand side of \eqref{eq:avaraged_langevin_equation_of_R_vector} is zero. In the equilibrium state (i.e.~when $\grad \vb*{u}$ vanishes), $\Gamma_{\mathrm{elast}}$ balances $\Gamma_{\mathrm{thermo}}$, leading to $\langle R^{2} \rangle \approx 3R_{\mathrm{eq}}^2$. For $\De \gtrsim 1$, since the thermal fluctuation $\Gamma_{\mathrm{thermo}}$ is negligibly small compared to $\Gamma_{\mathrm{fluid}}$ and $\Gamma_{\mathrm{elast}}$, \eqref{eq:avaraged_langevin_equation_of_R_vector} reduces to
\begin{gather}
    \Gamma_{\mathrm{fluid}} = -\Gamma_{\mathrm{elast}}.  \label{eq:balance_of_stretch_and_relaxation}
\end{gather}
This suggests that stretching by fluid is balanced with relaxation by polymers on average. Rewriting this equation as a timescale balance using $\langle R^2 \rangle$ yields
\begin{gather}
    \tau_{\mathrm{p, eff}} = \frac{\langle R^2\rangle}{\Gamma_{\mathrm{fluid}}} = - \frac{\langle R^2\rangle}{\Gamma_{\mathrm{elast}}},
    \label{eq:timescale_barance}
\end{gather}
which coincides with the effective relaxation time \eqref{eq:effective_polymer_relaxation_time}.

\FloatBarrier

\bibliographystyle{jfm}
\bibliography{jfm}

\end{document}